\documentclass[12pt]{article}
\usepackage{graphicx}
\usepackage{subfigure}
\usepackage{amsmath}
\DeclareMathOperator{\divergence}{div}
\DeclareMathOperator{\trace}{tr}
\DeclareMathOperator{\symm}{symm}
\newcommand{\vect}[1]{\mathbf{#1}}
\newcommand{\vectSym}[1]{\boldsymbol{#1}}
\newcommand{\tens}[1]{\mathbf{#1}}
\newcommand{\tensSym}[1]{\boldsymbol{#1}}

\newcommand{\fisher}{\tilde{\varphi}}

\newcommand{\valUnit}[2]{#1\,#2} 
\newcommand{\Watts}[1]{\valUnit{#1}{W}}
\newcommand{\perSec}{1/s}
\newcommand{\perK}{1/K}
\newcommand{\degK}{K}
\newcommand{\GPa}{GPa}
\newcommand{\GPaPerK}{GPa/K}
\newcommand{\TiFiveOneAlloy}{Ti-5Al-1Sn-1Zr-0.8Mo}
\newcommand{\TiFiveOne}{Ti5111}

\newcommand{\TiSixFour}{Ti64}

\begin{document}
%
%
\title{ A Methodology to Determine Tooling Interface Temperature and Traction Conditions
from Measured Force and Torque in Materials Processing Simulations
Based on Multimesh Error Estimation}
\date{}
\author{D.E. Boyce\thanks{Cornell University, Ithaca, NY },
        P.R. Dawson\thanks{Cornell University, Ithaca, NY (corresponding)} }
\maketitle

%
%
\begin{abstract}
  In this paper, we present a methodology for estimating average values for the
      temperature and the frictional traction over a tool-workpiece 
      interface using measured values of force and torque applied 
      to the tool.  
  The approach was developed specifically for friction stir welding and 
      friction stir processing applications, but is sufficiently general to be
      of use in a variety of other processes that involve sliding contact 
      and heating at a tool-workpiece interface.
  The methodology works with a finite element framework that is 
      intended to predict the evolution of the microstructural state  of the
      workpiece material as it undergoes a complex thermomechanical history
      imposed by the process tooling.  
  In our implementation, we employ a  finite element formulation is Eulerian 
      and three-dimensional; it includes
      coupling among the solutions for velocity, temperature 
      and material state evolution.
  A critical element of the methodology is a procedure to estimate 
      the tool interface traction and temperature from typical, 
      measured values of force and torque.
  The procedure leads naturally to an intuitive basis for estimating error 
      that is used in conjunction with multiple meshes to assure convergence.  
  The methodology is demonstrate for a suite of three experiments 
      that had been previously published as part of a 
     study on the effect of weld speed on friction stir welding of \TiFiveOne.
  The  probe interface temperatures and torques are estimated for
     all three weld speeds and the multi-mesh error estimation methodology 
     is employed to quantify the rate of convergence.
  Finally, comparison of computed and measured power usage is used 
     as a further validation.
  Using the converged results, trends in the material flow, temperature, 
      stress, deformation rate and material state with changing weld 
      conditions are examined .
\end{abstract}

  
%
%
%
%
\section{Introduction}
\label{sec:intro}
Friction stir welding (FSW) is a solid-state joining process in which frictional heating from
   the spinning probe softens the workpiece, thereby
   allowing the probe to advance steadily along an interface, mix the material and
   promote a metallurgical bond.
The process is characterized by pronounced interplay among the thermal and mechanical
  responses of the material and complex interactions between the probe and workpiece.   
The implications of these realities for developing and validating thermomechanical models are many.   
The solution methodologies should incorporate strong coupling of the equations that govern the 
   velocity, temperature and evolution of material state.  
The constitutive equations for the material behaviors must be accurate over the very broad ranges of 
   temperature and strain rate typical of FSW.   
The specification of the probe-workpiece boundary conditions should accommodate imprecise
   knowledge of the local conditions at the interface between the workpiece and probe. 
   
Similar complexities exist across a host of materials processing technologies
   and present a serious challenge to building numerical models that are  
    capable of supporting process development. 
In this regard, we expect that such a model will be able to accurately predict  
   how the material heats and deforms for particular process variables, 
   how this history alters the mechanical state,  and whether or not defects are 
   likely to occur.  
Reliability should be achieved through a program of model validation 
   that builds on documented knowledge of
   material behavior,  checks for internal consistency within a simulation, 
   and tests against observed trends. 
The model must be robust in the presence of uncertainties arising from 
  knowledge of the material properties or from uncertainties in the process conditions.
  
In this paper, we present a methodology for quantifying critical boundary conditions that
   arise in processing simulations, namely the frictional traction and interface temperature at interfaces between
   the tooling and the workpiece.   
Our approach draws on empirical measurements including  welding force, torque, 
   and total input power (derived from torque and force) 
   to estimate average values for the probe surface temperature and the frictional
   traction at the probe-workpiece interface.    
Specifying the probe boundary conditions in terms of temperature and traction, 
  in contrast to invoking heat flux and friction models,
  facilitates the direct use of process measurements to quantify conditions 
  at the probe-workpiece interface and to develop internal consistency checks
  of the simulation.
The process measurements can include a number of readily accessible macroscopic 
  resultants such as net forces, torques and thermal inputs. 
The methodology relies on a sophisticated material model that 
   includes thermal softening as well as strain hardening. 
We illustrate the robustness of our procedure by modeling experimental welding data 
   for a titanium alloy over multiple meshes.

The paper is organized as follows.  We first summarize each of the component parts of the simulation framework.  Namely, we describe  the modeling formulations for the flow field, the temperature distribution, and the material strength.  For each of these parts, we lay out the set of governing equations, the associated boundary and/or initial conditions, and the numerical solution method.  We then discuss the methodology for 
determining the probe temperature and traction for a particular set of process parameters from experimental data.  
This is followed by a summary of the experimental data for welding of a titanium alloy that we use to demonstrate the approach. Next, we present the simulations of these experiments and show how convergence on a set of boundary condition values is obtained.   From the converged results we then examine details of the flow fields, temperature distributions and strength profiles.    We conclude with a summary of the 
benefits of the approach.


\section{Modeling Framework}
\label{sec:sim}
\subsection{Background}
\label{sec:sim.model.background}

Finite element frameworks used to model materials processing take a variety of approaches in 
defining the reference frame for expressing the governing equations and establishing the computational domain.
Lagrangian, Eulerian, and ALE (Arbitrary Lagrangian/Eulerian) systems all have seen broad application.
(Consult \cite{bely_book}  for a general overview.)
The  work reported employs an Eulerian approach, principally because the deformations in some 
regions of the workpiece are quite large and using an Eulerian reference frame circumvents mesh distortion
issues in the context of very large strains.  
Using streamline methods in conjunction with an Eulerian framework facilitates
modeling the evolution of the material state during processing~\cite{daw_84,agr_daw_85,daw_87}.
Application of an Eulerian framework for modeling friction stir welding has been 
demonstrated previously by ourselves  
 in~\cite{cho_boy_daw_05,cho_boy_daw_2007}
as well as by others~\cite{Liechty2008Modeling}.

The current work is focused primarily on a robust procedure for specifying conditions at the
   probe-workpiece interface, but also includes the comparison of recovered surface 
   quantities with imposed quantities to estimate discretization errors.
The latter contribution falls broadly under the realm of \emph{a posteriori} error
    estimation.
Strictly speaking, the highly nonlinear models used in this work are beyond the
    scope of the present mathematical theory of error estimation for finite element methods, 
    but the approach is nonetheless useful.
For a general discussion of \emph{a posteriori} error
    estimation, see~\cite{Ainsworth1997Posteriori,Verfurth1996Review}.
The method addresses what has been lacking:   a systematic methodology to estimate
probe interface conditions from available data with a rigorous estimation of the related uncertainty.
Such a methodology is the focus of this paper.   It is combined with error estimation and verification of mesh convergence
to provide a framework for robust and reliable simulation of the coupled thermomechanical response of material during FSW.
\subsection{Eulerian Domain and Surface Specification}
\label{sec:sim.model.domain}

\begin{figure}[htpb]
  \begin{center}
    \includegraphics[width=4in]{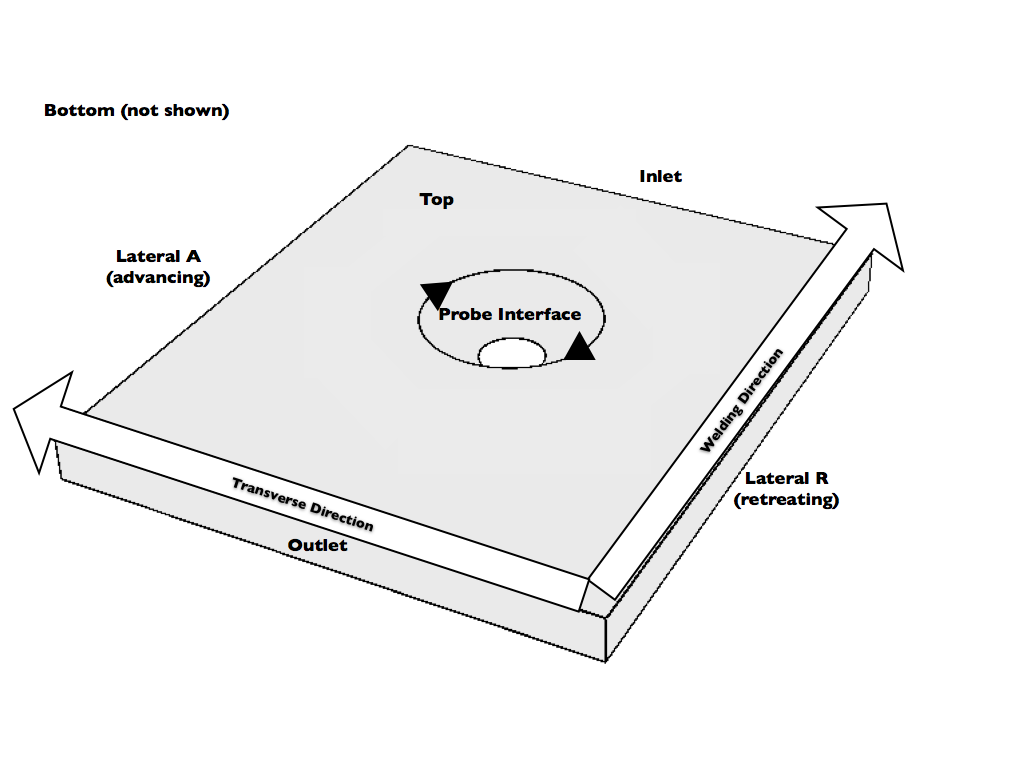}
    \caption{Surface designations for Eulerian domains used in the FSW simulations.}
    \label{fig:model-surfaces}
  \end{center}
\end{figure}

The computational domain of the model is a spatially-fixed control volume.
We are modeling the material flow relative to the probe, so the probe 
   is treated as fixed in space, and the workpiece is moving through the domain at the 
   designated weld speed.
Workpiece material enters from one end (inlet), 
   flows around and past the probe location,
   and exits (only) on the opposite end (outlet).  
The model geometry is shown in Figure~\ref{fig:model-surfaces}.
The surfaces are labeled for reference when describing
    boundary conditions.
The probe itself is not included in the simulation.  
Rather, the interface between the workpiece and 
    the probe interface is identified as a surface of the control volume 
    and boundary conditions are applied to it.  
Note that the direction of the probe relative to the workpiece, 
   typically referred to as the welding direction, 
   goes from outlet toward the inlet.
The probe is spinning, clockwise in the figure, which produces asymmetry
   on the two sides of the probe.
The \emph{advancing side} is on the left in the figure; on that side, 
   the probe motion due to spin is in the welding direction.
The \emph{retreating side} is on the right, 
  and the probe motion is in the reverse welding direction.
 %
%

\subsection{Governing Equations}
\label{sec:sim.model}

A system of governing equations is specified to represent the thermomechanical response of the 
workpiece as a steady-state process.  
We do not model the transient response during either the probe insertion or the probe removal.
The equations include balance laws for momentum, mass and energy, kinematic equations for the 
motion, constitutive equations for the material behavior, and boundary conditions. 
Altogether, the equations define spatial fields for the velocity, temperature  
   and material strength.
In the following subsections, we summarize the equations according to those used to determine each of the spatial fields 
   and the numerical formulations for solving the respective sets of equations.
The resulting systems are nonlinear and are 
    solved numerically using a fixed point iteration procedure.
The linear systems for each iterate of velocity and temperature 
    were solved in parallel using the PETSc~\cite{PETSc_2002}
    library.

\subsubsection{Equations governing the velocity field}
\label{sec:sim.model.mech}

Under the assumptions that viscous forces are much larger than inertial terms
and that body forces may be neglected, 
balance of linear momentum reduces to
\begin{equation}
  \label{eq:div-sigma}
  \divergence{\tensSym{\sigma}} = 0\\
\end{equation}
where ${\tensSym{\sigma}}$ is the Cauchy stress.   
From balance of angular momentum, the Cauchy stress is symmetric.
The deviatoric Cauchy stress is determined 
   from the deviatoric deformation rate, $\tens{D^\prime}$,
   through a viscoplastic constitutive law where 
   the underlying flow is assumed to be incompressible and elastic strains are neglected.  Under these assumptions
\begin{eqnarray*}
  \tensSym{\sigma}' &=& 2\mu \tens{D}' \\  
  \trace{\tens{D}} &=& 0
\end{eqnarray*}
The deformation rate is obtained 
from the velocity, $\vect{u}$, by
\begin{equation}
  \label{eq:kinematics}
  \tens{D} = \symm(\nabla \vect{u})
\end{equation}
The effective viscosity, $\mu$ , is inferred from the relations for the rate-dependent flow stress of the material.  
We employ an empirical equation for the flow stress suggested by Kocks and Mecking~\cite{kocks_mecking_2003} 
\begin{equation}
  \label{eq:constitutive-law}
  \bar{\sigma} = \tau(s,\theta) f(\theta)  
           {\biggl(\frac{\bar{D}}{D_0}}\biggr)^{m(\theta)}  
\end{equation}
In this equation, the  effective stress, $\bar{\sigma}$, is related to the effective
    deformation rate, $\bar{D}$, through a power-law dependence in which
    the  deformation rate is normalized by a reference deformation rate, $D_0$.
    $\theta$ is the temperature.
The exponent in the power-law relation is $m$, and is referred to as the rate-sensitivity.    
The effective stress and deformation rate are computed from the corresponding tensors 
   according to  $\bar D^2 = \tens{D}':\tens{D}'$ 
   and $\bar\sigma^2 = \tensSym{\sigma}':\tensSym{\sigma}'$.
There are two scaling terms for the flow stress: $\tau$ and $f$.
$\tau$ is a function of the state variable for 
    strength, $s$, and the temperature-dependent shear modulus, $G$.
 $f$ is a  temperature-dependent term characterizing the activation energy. 
The specific forms of $\tau$ and $f$ are
\begin{eqnarray*}
  \tau(s,\theta) &=& s\,G(\theta) \\
  f(\theta)     &=& \exp\biggl(   \frac{Q}{R} \bigl(  \frac{1}{\theta} - \frac{1}{\theta_0}\bigr) \biggr)
\end{eqnarray*}
where $Q$ is the
activation energy, $R$ is the universal gas constant, and $\theta_0$ is a reference temperature.  
Parameters $m$ and $G$ are linear functions of temperature
\begin{eqnarray*}
  m(\theta) &=& m_0\,+ m_t \theta \\
  G(\theta) &=& G_0\,+ G_t \theta
\end{eqnarray*}
This model has two features that are particularly important to dealing with the wide range
of strain rates and temperatures characteristic of friction stir welding simulations:  the scaling
of the flow stress by the temperature-dependent shear modulus and the explicit control of the
rate sensitivity through the power law.  

Boundary conditions are either specified velocity
   or traction components on each surface.
The material contacting the probe is constrained to have
    zero velocity normal to the probe surface and  a frictional
    traction that is applied tangentially to produce a desired torque.
The weld speed is specified by specifying velocity of material 
    entering the control volume at the inlet.
Motions on the top, bottom and the two sides are also constrained
    to be tangential to those surfaces, respectively. 
Finally, the outlet has a zero traction everywhere.

\subsubsection{Equations governing the temperature field}

Balance of energy provides the governing equation for the temperature field, and may be written as
\label{sec:sim.model.temp}
\begin{equation}
  \label{eq:div-q}
  -\divergence (\kappa\nabla\theta) +  \rho c_p \nabla\theta \cdot \vect{u}=  \dot{\mathcal{Q}}
\end{equation}
The first term quantifies heat transfer by conduction  while the second quantifies advection.   $\dot{\mathcal{Q}}$ is the heat generated in the body as a consequence
of the viscous dissipation associated with the deformations ($\dot{\mathcal{Q}} =\tensSym{\sigma}':\tens{D}'$).
Material constants are the conductivity, $\kappa$, the mass density,  $\rho$, 
   and the heat capacity, $c_p$.  
All of these properties are functions of temperature.   

Boundary  conditions for the temperature field can be either 
   a specified temperature or a heat flux.
Several types of flux conditions are used.  
For convection and surface contact, the heat loss through a surface
    is linearly proportional to the difference in surface and ambient temperatures.
For radiation, 
   the flux is proportional, through the emissivity, to the fourth power of 
   the difference in surface and far-field temperatures.
The temperature is specified for the material entering the control volume at the inlet.
The probe temperature is also specified using a value determined using the
    procedure detailed in Section~\ref{sec:bc_method}.
The top surface has a radiation condition, while the bottom surface 
    has a heat flux due to contact with the anvil.
The outlet has a zero (spatial) flux condition, 
    but note that there is heat transfer across that boundary 
    due to advection of mass exiting the control volume.

\subsubsection{Equations governing the evolution of the state variable}
\label{sec:sim.model.state}

The state variable, $s$, quantifies the  strength and is a key part of the constitutive
   model discussed above.
Again, we employ a model proposed by Kocks and Mecking~\cite{kocks_mecking_2003}.
Although this model was developed for face-center cubic polycrystalline materials, our
experience is that it also works well for titanium, as well.
The state variable is associated with a differential volume of material that may be taken as
a point in the computational domain and evolves as a consequence of the
deformation according to 
   \begin{equation}
     \label{eq:s-dot}
  \frac{Ds}{Dt} = \frac{h_s}{G(\theta)} (1 - \frac{s}{s_s})^{n_s}\bar{D}\\
\end{equation}  
  where
\begin{eqnarray}
  \label{eq:s-sat} 
   s_{s}  &=& {\Biggl( a_s + b_s \biggl(\frac{{\tilde{\varphi}(\theta, \bar{D})}}{{\tilde{\varphi_s}}}\biggr)^{\frac{1}{2}}\Biggr)^2 }\\
    \label{eq:phi}
  \tilde\varphi(\theta, \bar{D}) &=& \frac{\theta}{G(\theta)} \, 
        ln\, \biggl(\frac{\mathit{D_s}}{\bar{D}}\biggr)\\
    \label{eq:phi-ref}
  \tilde\varphi_s  &=& {\tilde\varphi}(\theta_r, \mathit{D_r})
\end{eqnarray}
Equation~\ref{eq:s-dot} gives the material derivative of $s$ in
   terms of the shear modulus,  the effective deformation rate, the state saturation, $s_s$,
   and material parameters, $h_s$ and $n_s$.
The state saturation is given in Equation~\ref{eq:s-sat} as a 
   function of the  Fisher factor, $\fisher$ (Equation~\ref{eq:phi}).
The Fisher factor is     
   normalized by a reference value (Equation~\ref{eq:phi-ref}),  namely the 
   Fisher factor at a reference temperature and deformation rate, $\theta_r$ and $\mathit{D_r}$.
The constants $a_s$ and $b_s$ are material parameters.
For FSW applications, a critical feature of the evolution equation is the scaling of the 
   state saturation with deformation rate and temperature.  It is through the state saturation
   that bounds  are imposed on the increases in flow stress from strain hardening/softening.  Bounding the flow stress 
   is central to maintaining realistic values of the computed stress and temperature over the 
   large excursions of temperature and strain experienced in FSW.

For the state evolution, initial conditions are specified 
    on the inlet surface.
Equation~\ref{eq:s-dot} is integrated along  streamlines of the flow field  to
determine how the state of material 
evolves as it passes through the control volume.  
For flows with recirculations (closed streamlines within the control volume), such as can occur on the probe surface, 
    special treatment must be given.
For points on a closed streamline,  we assign the saturation value to the material state,
as defined by Equation~\ref{eq:s-sat}.

\subsection{Numerical solution of the governing equations}
\label{sec:sim.model.fem}

The numerical solution of the governing equations 
   proceeds as in~\cite{cho_boy_daw_2007}.
   For completeness, we include a brief statement of 
   methodology used for each of the primary field variables (velocity, temperature and state variable).

The velocity field is determined from a weak form of the equilibrium equation: 
\begin{equation}
  -\int_V \tensSym{\sigma} : \nabla\,\tensSym{\Psi} \,{\rm dV}
  +\int_S \vect{t}\cdot \vectSym{\Psi}\,\mathit{dS}\,=\,0
\label{eq:virtualw}
\end{equation}
where $\vect{\Psi}$ is a vector weighting function, 
    $\mathit{V}$ is the workpiece volume, 
   $\mathit{S}$ is its surface, and  $\vect{t}$
   is the surface traction. 
The viscoplastic constitutive equations for the flow stress are used 
   to express the stress in terms
   of the deformation rate. 
The  incompressibility constraint is enforced by  
   a weighted residual as: 
\begin{equation}
       \int_V\,{\psi}\, \divergence\vect{u}\, dV\,=\, 
       \int_V\,{\psi}\, {\trace(\tens{D})}\, dV\,=\,0
    \label{eq:weighted_residual}
\end{equation}
where ${\psi}$ is a scalar weighting function.  
This equation acts as a constraint on the motion allowed by Equation~\ref{eq:virtualw}
   and is treated by the consistent penalty method~\cite{engelman_1982,lee_1989}.

To determine the temperature field, a residual is formed using the energy equation. 
Using the scalar weights, $\psi$, the residual may be written as: 
\begin{equation}
   \int_V\,[\divergence(\kappa\nabla\theta)
    -\rho c_p \vect{u}\,\cdot\, \nabla\theta\,+\dot{\mathcal{Q}}]
    \,\psi\,\mathit{dV}\,=\,0.
\label{eq:gdiscrete}
\end{equation}
Streamline upwinding~\cite{book_fem2000,brooks:supg_1982}
   was used to stabilize the solution against spurious spatial oscillations.

We use the streamline integration method to solve for the material state
   variable at the nodal points.  For steady flows, the material derivative 
   in Equation~\ref{eq:s-dot} retains only the convective term
   \begin{equation}
 \frac{Ds}{Dt}= \vect{u} \cdot \nabla s 
\label{eq:streamline}
\end{equation}
This allows the evolution equations for the state variable
   to be written along characteristic lines of the partial differential equation
    and integrated point-by-point as ordinary differential equations.
 To obtain the value of the state variable at a given node of the finite element mesh, 
   we first trace the streamline passing through that node upstream to 
   a point within the mesh where the state has already been evaluated.  
We use the value at that point as an initial condition for the 
    evolution equation and then integrate forward to determine
    the value at the starting node.
By progressing from the inlet to the outlet through the mesh, 
   the  distribution of state variable over the 
   complete mesh may be determined efficiently~\cite{daw_84,agr_daw_85,daw_87}.
%


\section{Determining the Probe Boundary Conditions }
\label{sec:bc_method}
To solve the governing equations, 
   we need to specify values for the boundary conditions.  
For most surfaces on the workpiece, this is straightforward.  
However, values for the 
    traction and temperature applied to the workpiece at the interface 
    with the probe are problematic.   
Values for these quantities are particularly difficult to ascertain 
    as this interface  is not accessible to instrumentation.
Here we present a methodology for estimating the local
   probe/workpiece interface conditions from experimentally known quantities 
   associated with the probe, namely its 
   translational and rotation rates and the resultant weld (probe) force and torque.
Corresponding quantities can be computed from simulation data 
   for specified tractions and temperature
   combinations applied to the workpiece/probe interface.

The methodology consists of  adjusting the traction and temperature  
   in a systematic way so that the simulated
   probe quantities come into agreement with measured values.  
To accomplish this, we make two simplifying assumptions regarding
   the distribution of frictional traction and interface temperature:  both are uniform
   over the probe surface.  Specifically, 
the applied traction is of constant magnitude and in the direction
    of rotation; it can derived directly from the measured torque.
The applied temperature is uniform over the probe surface;
    its value can be derived indirectly from the weld force, given an accurate 
    characterization of the dependence of the flow stress on strain rate and temperature.
An iterative method is needed to identify the value of temperature for a 
    particular mesh.    
The methodology delivers average
    interfacial frictional tractions and temperatures
    with resultants equal to those measured on the probe.   

The simplicity of the boundary conditions enables a 
    straight-forward check on the adequacy of the mesh to capture the true solution.
The check is based on a comparison of the applied and recovered torques in
    the simulations.
By repeating the simulations on meshes
    of increasing resolution,  it is possible to confirm that the 
    discretization error tends to zero and the solution converges to within an 
    acceptable tolerance.
We consider this error check to be a powerful tool in its own
    right and a central feature of our methodology in general. 
In addition, the error analysis  revealed a key relationship 
    between recovered values of torque and weld force
    in the FSW simulations, which we were able to exploit effectively 
    in comparing simulation with experiments.    
Additional validation, independent of the values of probe traction and temperature,
   is provided by comparison of thermal fluxes
    with measured power consumption. 
Details are given below and in Section~\ref{sec:applic}.

\subsection{Resultants of Force, Torque, Heat Flux and Power}
\label{sec:bc_method.resultants}

The resultant quantities of interest include the resultant forces, torques, 
    heat fluxes and power.
These are defined analytically as integrals of appropriate
    quantities over a particular surface or volume of interest.
For a given finite element mesh, these integrals 
    are integrated numerically using the approximate solution.
The resultant force on a surface is the integral of the traction:
    \mbox{$\int_S \vect{t} \:dS$}, where $\vect{t}$ is the traction
    on surface $S$.  
When computed over the
   inlet surface, the resultant is the reaction needed to maintain equilibrium 
   when the probe force is applied to the workpiece, and thus it
   corresponds to the \emph{weld force}.
The mechanical power on a surface due to a traction
   acting on material moving along the surface is computed by 
   \mbox{$\int_S \vect{t} \cdot \vect{u} \:dS$}, 
   where $\vect{u}$ is the velocity field on the surface; for the
   inlet surface, in particular,  this is the \emph{weld power}.
The net heat flux crossing a surface is computed as 
   $\int_S \vect{q} \cdot \vect{n} \;dS$,
   where $\vect{q}$ is the heat flux and $\vect{n}$ is the surface normal.
If $\dot{\mathcal{Q}}$ is the heat generation density due to deformation,
   then the total heat generation rate over the body 
   $V$ from mechanical dissipation is given by 
   $ \int_V  \dot{\mathcal{Q}}\;dV$.

The torque calculation is of particular interest.
The torque is the moment about the probe's axis of rotation 
   caused by the frictional tractions.
The moment is computed as the integral over the probe surface 
   of the vector cross product between the frictional traction $\vect{t}^f$ 
   and its associated moment arm about the probe axis.  
We represent the probe axis of rotation using a point $\vect{p}$ 
    on the axis and a (unit vector) direction  $\vect{a}$.   
For a point $\vect{x}$ on the surface,
     the moment arm $\vect{r}$ is the vector connecting the
     axis at point $\vect{p}$ to the point of application:
      \mbox{ $\vect{r}$  = $\vect{x} - \vect{p}$}.
To obtain the torque  about the rotation axis, 
     we take the component of the moment aligned with $\vect{a}$, 
     resulting in a scalar value for the torque.
Thus, the resultant torque  $T$ about the probe axis of rotation 
     from the action of the frictional tractions on the probe surface 
     is given by
\begin{equation}
  \label{eq:torque}
  T = \vect{a} \cdot \int_S \vect{r} \times \vect{t}^f \;dS
\end{equation}

\subsection{Estimation using force and torque resultants}
\label{sec:bc_method.procedure}
The full procedure for determine boundary values to 
    apply on the probe interface involves solving
    the finite element systems on two or more meshes.
By proceeding from coarser to finer meshes, 
	it is possible to estimate resultants on more finely resolved
	meshes from analysis of error associated with individual meshes.    
In this section, we describe the complete process beginning
	with the determination of the traction and temperature for a single, fixed mesh.

\subsubsection{Traction and temperature for a fixed mesh}
\label{sec:bc_method.single}
The basic procedure of the methodology is to find
    a traction and a temperature for the
    probe interface that will give 
    resultants for probe torque and weld force that are equal to the measured values.
The traction value is determined by computing the torque assuming  a unit
    traction in Equation~\ref{eq:torque} and rescaling it to match the desired 
    value of torque, \emph{i.e.} the measurement.
The temperature is found indirectly by determining the probe temperature that leads to
    a computed weld force that is equal to the measured weld force.
This procedure involves finding the solutions for several possible probe temperatures
    and from these estimating the temperature that will provide the best
    match between measured and computed weld force.    
We note here that the weld force that is recovered from each solution 
   will be adjusted based on the estimated error (described below), and that
    adjusted value will be used for comparison with the measured 
    weld force.
 In the FSW system, 
    when the assumed probe temperature increases, the recovered
    weld force decreases.
By solving the FEM equations with different assumed probe temperatures
    and computing the associated weld forces, one can quickly
    narrow the range of probe temperatures that provide reasonable weld forces.
Our experience is that this iteration converges quickly
    and only a handful of systems need to be solved.
In the end, we obtain a temperature and a traction to apply
   on the probe interface which are consistent the measured
   torque and weld force.
    
\subsubsection{Estimation of error for a fixed mesh}
\label{sec:bc_method.errors}

The procedure outlined in Section~\ref{sec:bc_method.single} provides values for probe traction and temperature 
for which the simulation results match target values
for torque and force for a specified mesh.
This process does guarantee that the solution using that mesh is otherwise accurate, however.  
To make that assessment the discretization error associated with the mesh must be estimated.
Because we use a uniform traction condition on the 
    probe interface,
   we have a natural check on the quality of our solution.
After a solution is computed for a given mesh and set of boundary conditions, it is postprocessed to 
   compute macroscopic resultants, including the torque.
In this computation, the internal stresses are evaluated at the points on the boundary from the deformation rate
and temperature using the constitutive equations.  If the solution estimates either of these poorly at critical points, then the stress will be
poorly estimated at those points as well.    The consequence here is that the torque recovered from the internal stress may not match
the torque applied externally via the boundary conditions.    Thus, we use a comparison of the recovered torque
to the applied torque to determine the magnitude of the discretization error in the deformation rate and temperature
and the need for additional mesh refinement.  

Additionally, for the particular models used in this paper,
    we found and exploited a relationship between the 
    error in the torque and the recovered weld force.
The recovered torque was too low in each case, with the finer results
    being closer to, but still below, the target value.
Meanwhile, the recovered weld force decreased with finer mesh resolution,
meaning that the estimate for weld force from any given mesh was consistently too high.
This trend applied to all the cases we considered in the application presented in Section~\ref{sec:applic}.
The error in the recovered torque and the error in the estimate of the weld force correlate 
as both stem from the discretization of the workpiece.  Consequently, 
it is possible to compute a better estimate of the weld force by adjusting
the computed weld force using a factor proportional to the torque error.  
Equation~\ref{eq:adj} gives the formula for the
    weld force adjustment,
    where $f_{\rm adj}$ is the \emph{adjusted weld force}, 
    $E$ is the relative (proportional) error in the torque,
    and $f_{\rm rec}$ is the recovered resultant weld force from the simulation
\begin{equation}
  \label{eq:adj}
  f_{\rm adj} = (1 - E) f_{\rm rec}
\end{equation}
The adjustment to the computed weld force removes most of the
    effect of mesh size, providing a consistent value across
    meshes.
This is what we use in our comparisons in Section~\ref{sec:applic}.

\subsubsection{Reduction of error by mesh refinement}
\label{sec:bc_method.resolution}
The procedure for a single fixed mesh is repeated on meshes 
   with greater resolution until the solution has converged to within a specified tolerance.
The discrepancy between the torque recovered by postprocessing and the torque used to compute the applied friction traction 
   serves as a effective measure of the convergence of the entire solution.
By repeating the procedure described above on different meshes,
   one can check that the solution is converging by monitoring the discrepancy in torques.
To make this quantitative, we need to prescribe a measure of the 
   element size, commonly referred to using the variable $h$.
Here, we use $h = \sqrt[3]{1/n}$,  where $n$ is the number of nodes 
   in the mesh.
While our meshes are not perfectly regular, they are based on
   sections with regular subdivisions, so this is roughly proportional
   to the element size.
In section~\ref{sec:applic}, our errors in traction will be seen to have 
   a very strong linear convergence with $h$.

\subsection{Validation using power resultants}
\label{sec:bc_method.thermal}

Finally, other measured resultants should be compared
   with the recovered values.
The thermal resultants are of particular interest
   because we used the mechanical resultants to calibrate
   the probe traction and temperature.
One resultant we have available for comparison is the total power used to rotate the probe.
  which is computed from the product of the rotation rate and the measured torque.
However, we do not know the partitioning of the power between the workpiece,
     the tooling, and the environment.
Still, the total power provides an upper bound on the amount of heat generated.
For our application, the solution approaches this limit with increasing
     translational speed, indicating a greater fraction of the total power
     is transferred to the workpiece as the probe translation rate increases.

%
\section{Application to FSW of Ti5111}
\label{sec:applic}

We illustrate the methodology to determine probe interface boundary 
   conditions from knowledge of the force and torque applied to the
   probe via the tooling  
   on an example of friction stir welding of a titanium alloy.
Process parameters include weld speed,  rotation speed, probe and 
   workpiece geometries, as well as material properties. 
 The mechanical properties are taken from published values or
   from fitting of model constants to published data.

\subsection{Experimental Data}
\label{sec:applic.data}
The process data comes from FSW experiments described 
in greater detail in~\cite{wolk:thesis} and summarized briefly here.
Three welds were performed on quarter-inch sheet made of the 
   titanium alloy, \TiFiveOneAlloy\ (\TiFiveOne).
All three welds were made at the same probe rotational speed of \valUnit{225}{rpm}, but at different weld speeds:
   a slow speed of 1.0 ipm (inches/minute), 
   a medium speed of 2.5 ipm, and a fast speed of 4.0 ipm.
The data include weld force profiles, from which we took 
   nominal values of 31 kN, 36 kN and 38 kN for the slow, medium and
   fast weld speeds, respectively.
The torque was  not measured independently for all weld speeds,
but a nominal value of 47 Nm was provided for all three cases. 
As a check, we can compute the rate of power input from the torque and
   rotation speed, giving approximately 1,100 W; the power contribution from the 
   forward movement of the probe is much less, but increases with weld speed.
This data set lacks detailed information on 
   heat loss through the probe via controlled cooling and so 
    only a partial check on the partitioning of input power is possible.
Microstructures and textures for these welds are examined in~\cite{Fonda2010Texture}.
   
Parameters for the thermal and mechanical constitutive models summarized in Section~\ref{sec:sim}
  were estimated from available data in the literature~\cite{asm_handbook}.
For the Kocks-Mecking model, the parameters were found by fitting to data
    from the \cite{AtlasOfFormability}  for \TiSixFour.
These data were used as a reasonable representation of the \TiFiveOne\ behavior
   in the absence of adequate data for \TiFiveOne\ itself.    
Values of the model parameters
    are given in Table~\ref{tab:matparm}.
In~\cite{Pilchak2011Microstructure}, experimental thermocouple data for \TiSixFour, 
   indicated that stir zone temperatures exceed the beta transus. 
To allow for the change of phase,
   we use two sets of parameters for the two regimes.  
One set applies to the higher temperature BCC phase, the other set
   to the lower temperature HCP phase.
The thermal properties, however, were taken as the same for both phases.
%
%
\begin{table}
    \caption{Material parameters used in simulations.}
    \label{tab:matparm}
    \centering
    \begin{tabular}{|c|c|c|c|} \hline
	\textbf{parameter}& \textbf{unit} & 
        \textbf{phase 1}  & \textbf{phase 2}  \\ \hline
        \multicolumn{4}{|c|}{\textit{Flow Properties}} \\ \hline
                $G_0$      &  \GPa    &   43.4   & 43.4    \\ \hline
                $G_t$      &  \GPaPerK&   -0.021 & -0.021    \\ \hline
                $m_0$      &  -       &  0.05    & -0.685   \\ \hline
                $m_t$      & \perK    &  0.0     & 0.0007   \\ \hline
                $Q/R$      & \degK    &  100.0   & 1000    \\ \hline
                $\theta_0$ & \degK    &   462.0  & 900.0    \\ \hline
                $D_0$      & \perSec  &   1.0    & 1.0    \\ \hline
        \multicolumn{4}{|c|}{\textit{State Evolution}} \\ \hline
                $a_s$      &  - &  0.17  & 0.304  \\ \hline
                $b_s$      &  - &  -0.02   & -0.185     \\ \hline
                $h_s$      &  \GPa  &  50.0   & 20.0 \\ \hline
                $n_s$      &  - &  1.0   &  1.0 \\ \hline
                $D_s$      & \perSec   &  $10^{7}$   & $10^{7}$ \\ \hline
                $\theta_r$  & \degK      &   1150   & 1150   \\ \hline
                $D_r$       & \perSec    &   1.0    & 1.0  \\ \hline
        \multicolumn{4}{|c|}{\textit{Thermal Properties}} \\ \hline
            $\rho$     &kg/$\mbox{\textrm{m}}^3$  & \multicolumn{2}{c|}{4430}    \\  \hline
            $c_p$      &J/kg K                    &  \multicolumn{2}{c|}{533}  \\  \hline
            $\kappa$   &W/mK                      &  \multicolumn{2}{c|}{7.5}    \\ \hline
    \end{tabular}
\end{table}

\begin{figure}[htpb]
  \begin{center}
   \subfigure[]{\includegraphics[width=2.3in]{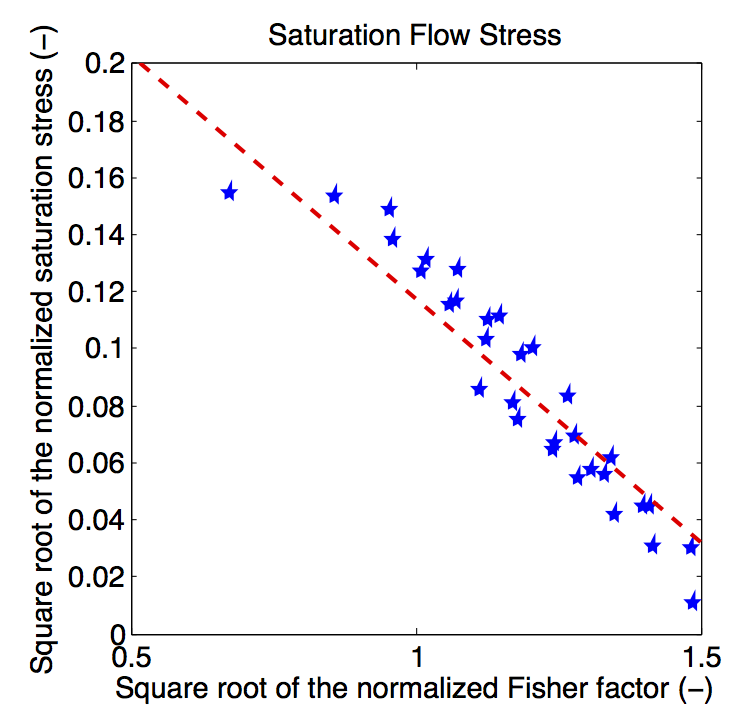}}
   \subfigure[]{\includegraphics[width=2.7in]{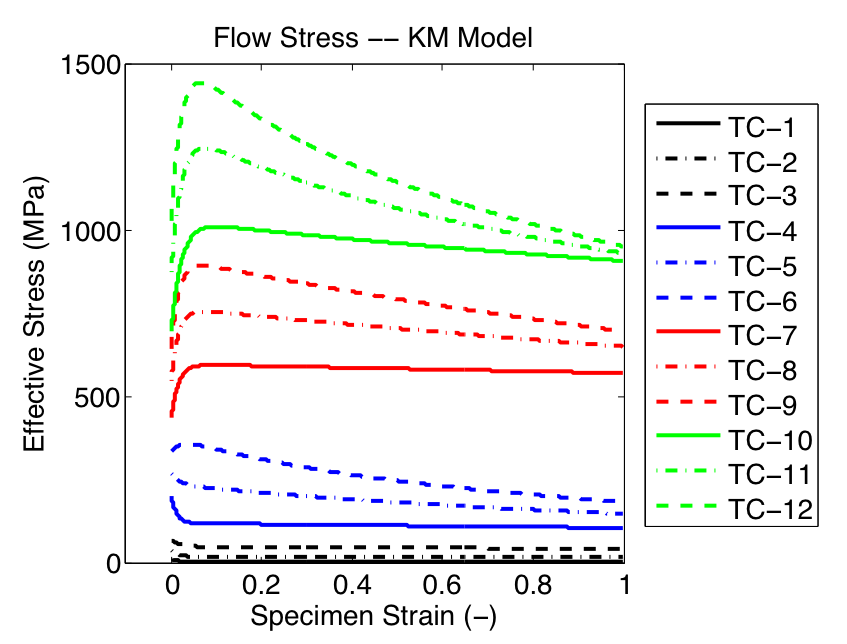}}

   \caption{Computed stress strain curves for polycrystalline \TiFiveOne\ 
   using the  Kocks-Mecking model.
   Caption labels indicate test strain rate and temperature $(\bar D (1/s),\theta (K))$ according to: TC-1=(0.01,1273);
   TC-2=(1.0,1273);
   TC-3=(25,1273);
   TC-4=(0.01,1073);
   TC-5=(1.0,1073);
   TC-6=(25,1073);
   TC-7=(0.01,673);
   TC-8=(1.0,673);
   TC-9=(25,673);    
   TC-10=(0.01,273);
   TC-11=(1.0,273);
   TC-12=(25,273);  }
    \label{fig:sat-stress}
  \end{center}
\end{figure}

As discussed in Section~\ref{sec:sim}, of particular interest is the saturation stress,  which appears in the 
    evolution equation for the strength and is a function
    of temperature and strain rate.  
The saturation stress bounds  the flow stress  in the limit
    of large strain deformations and serves to prevent the computed
    stresses from becoming unrealistically high as a consequence of the large strains 
    imposed on the material as it passes close to the probe.    
The saturation stress as a function of normalized Fisher factor is shown in Figure~\ref{fig:sat-stress}
    based on flow stress data reported for \TiSixFour~\cite{AtlasOfFormability}. 
Also shown are the computed stress-strain histories for simulated tensile tests corresponding to 
   a range of strain-rate and temperature combinations.   Note that under the assumed starting state,
   many of the responses demonstrate softening (a reduction of flow stress) as deformation
   heating raises the material temperature, which in turn lowers the saturation stress.
%
%

\subsection{Simulation Geometry}
\label{sec:applic.mesh}
\begin{figure}[htpb]
  \begin{center}
    \includegraphics[width=2.5in]{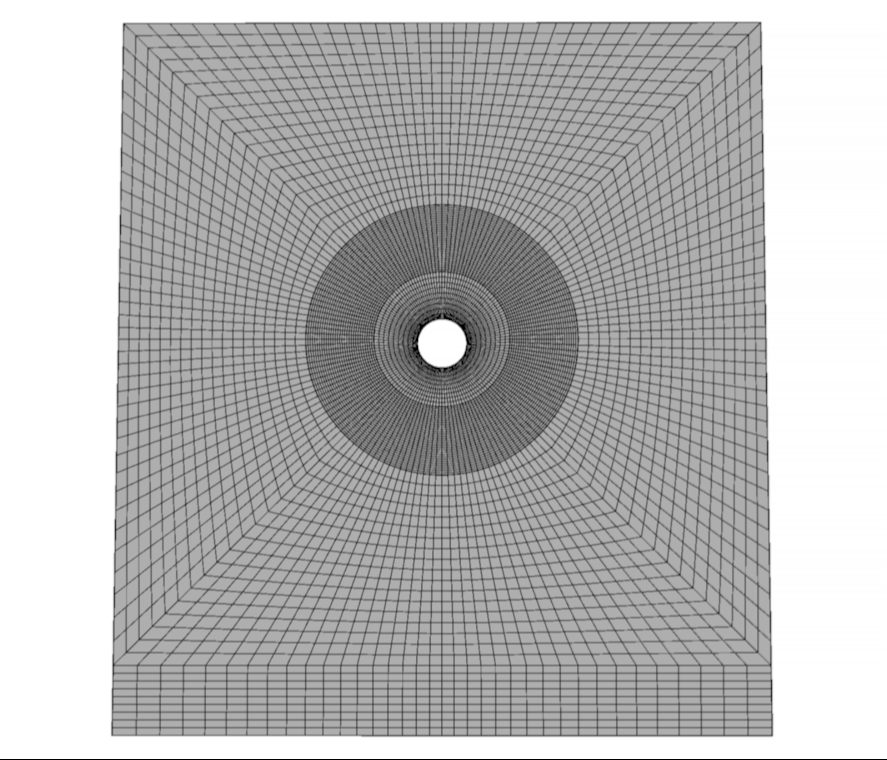}
    \includegraphics[width=2.6in]{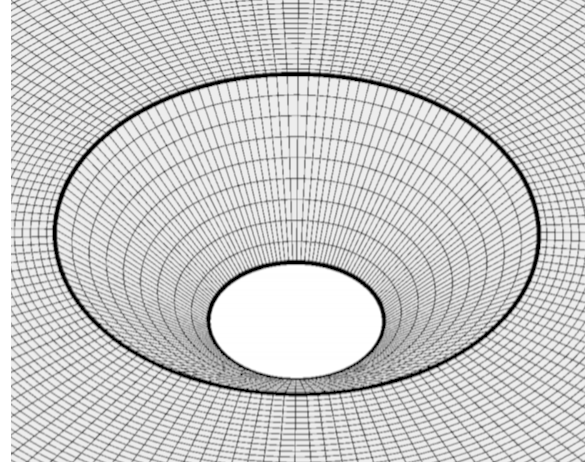}
    \caption{Simulation geometry: (left) view from above, 
      (right) close-up view from above of the probe surface discretization.}
    \label{fig:mesh}
  \end{center}
\end{figure}

The model geometry from Figure~\ref{fig:model-surfaces} 
     is shown in Figure~\ref{fig:mesh} with the finest mesh discretization used in the simulations discussed here.
The domain width was  \valUnit{120.2}{mm};
   the length of the region ahead of the probe was \valUnit{60.1}{mm},
   and the length behind the probe was  \valUnit{72.75}{mm}.
The domain is extended downstream of the pin (one additional
   probe length) to better capture the lateral gradients in the temperature field.
The probe shape was a truncated cone with  
   diameter of \valUnit{12.65}{mm} at the top 
   and \valUnit{4.65}{mm} at the bottom.
No shoulder is used in this model, consistent with the experimental conditions.
%
%
Three meshes, using 20-node brick elements, were used.
Mesh A was the coarsest, having 5,376 elements and 26,488 nodes.
Mesh B was a direct refinement of Mesh~A, having twice
    the resolution in every direction; it had 43,008 elements and 191,696 nodes.
Mesh C had the finest resolution and was used for the final results;
     it had 65,124 elements and 286,641 nodes.

\subsection{Probe Interface Boundary Conditions}
\label{sec:applic.bcs}

\begin{figure}[htpb]
  \begin{center}
    \includegraphics[width=2.5in]{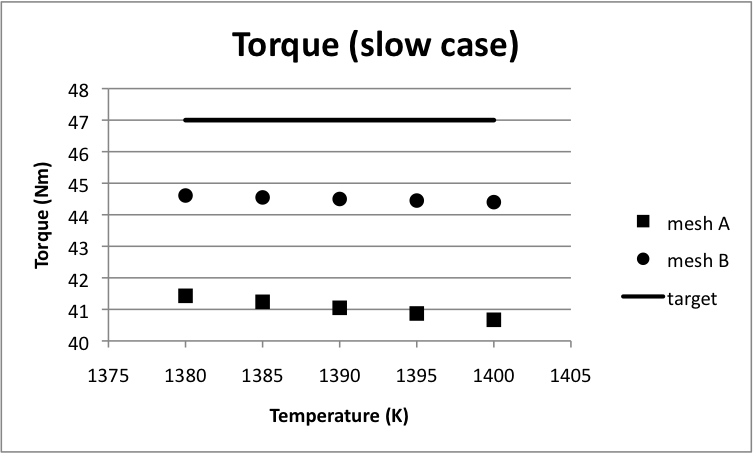}
    \includegraphics[width=2.5in]{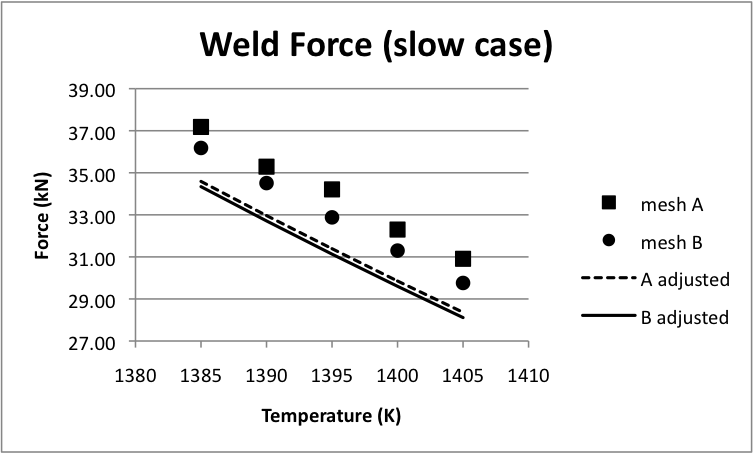}
    \caption{Relation of torque error to computed weld force
for two mesh resolutions, including adjusted weld force.}
    \label{fig:extrap}
  \end{center}
\end{figure}

The  procedure to determine boundary values for the probe interface
     was described in Section~\ref{sec:bc_method}.
We applied this procedure for each of the three experimental cases with different 
    weld speeds. 
 The methodology begins with estimation of the probe traction and temperature
 on the coarsest mesh, Mesh~A.     
The probe traction is computed directly from the known value of the torque
   using Equation~\ref{eq:torque}.  
The measured data indicated the torque was approximately the same for
    all three weld speeds. 
For the probe geometry used here, a probe traction of \valUnit{9.23}{MPa}
   produces the measured torque value of \valUnit{47}{Nm}.
Using this traction for all three weld speeds, we ran several simulations spanning a range of 
   assumed probe temperatures.   
The simulation results were postprocessed to obtain the various resultant
    quantities of interest, with the recovered torque and force for the slowest weld speed
    shown in Figure~\ref{fig:extrap}.
 The torque and force both decrease with probe temperature over the range
     in probe temperatures from 1380K to 1400K (for the slowest case).  
 For the torque, the error in the recovered torque ranges from 10\% in the
    slow case to nearly 40\% in the fast case.
 As described in Section~\ref{sec:bc_method},
    the recovered weld forces were adjusted based on the
    error in the recovered torque.  
Based on the trends illustrated in Figure~\ref{fig:extrap},
   the target weld force  of \valUnit{31}{kN} is achieved using a probe  temperature
   of \valUnit{1390}{K} for the slowest weld speed of 1 ipm.   
The probe temperatures for the 2.5 and 4 ipm for  Mesh~A were
   \valUnit{1470}{K} and \valUnit{1530}{K}, respectively.
   
To reduce the discretization error, the  simulations were repeated with Mesh~B, which has twice
   the resolution of Mesh~A in all  directions,  using the same input parameters.
For comparison, the recovered resultants for this mesh are also shown on Figure~\ref{fig:extrap}.
The recovered torque and force
   both move closer to their respective target values.  
The torque errors of Mesh~B were very nearly half those
   of Mesh~A, indicating a linear convergence with mesh size.
Note also that the adjusted weld forces for the two meshes are nearly the
   same over the entire range of probe interface temperature.
This indicates that the adjustment, based on the estimated error in torque, 
   removes a primary effect of mesh size, allowing a consistent 
   estimation of weld force independent of the mesh.
A final simulation was done for Mesh~C at the optimal temperature
  suggested by the analyses on Meshes~A and B.
After postprocessing to obtain the resultants,
  the same adjusted weld force was obtained.
     
\begin{figure}[htpb]
  \begin{center}
    \includegraphics[width=4.0in]{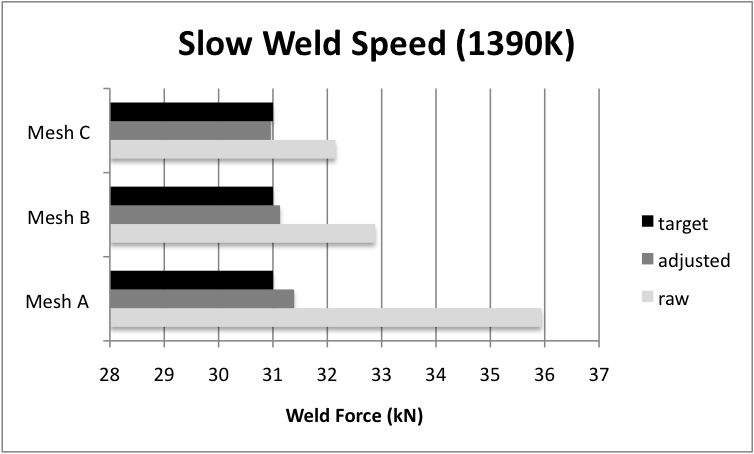}
    \includegraphics[width=4.0in]{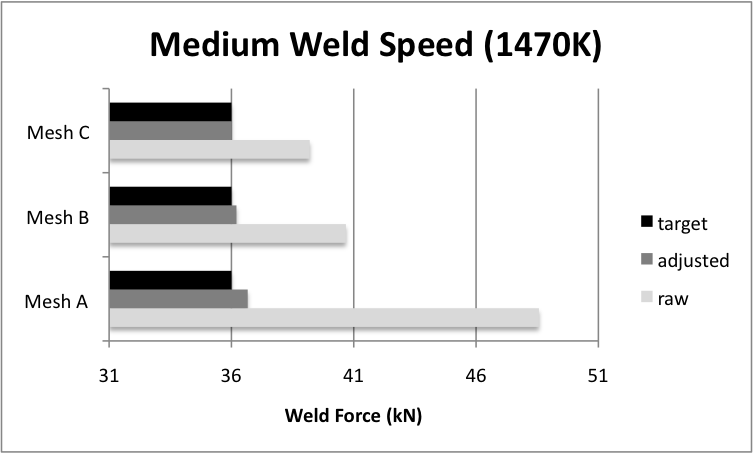}
    \includegraphics[width=4.0in]{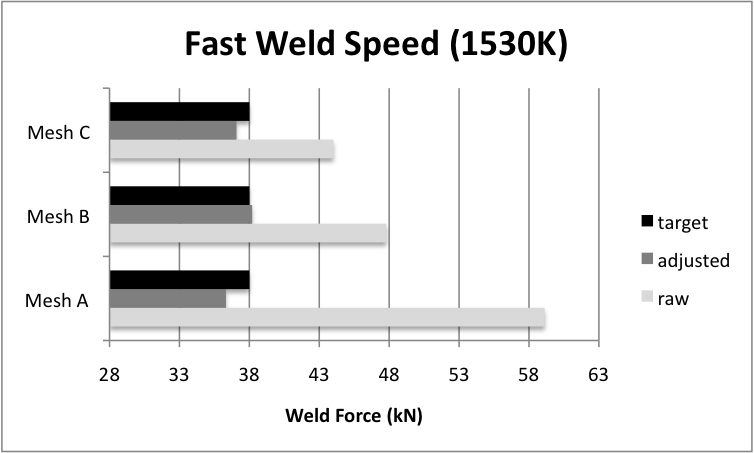}
    \caption{Weld force results at optimal temperatures for all weld speeds.
Plots show recovered, adjusted and target weld forces for each of three meshes.}
    \label{fig:best-results}
  \end{center}
\end{figure}

 Figure~\ref{fig:best-results} summarizes the weld force results at the
    optimal temperature for all three weld speeds and all three meshes.
 Mesh results are shown from top to bottom, corresponding to 
    finest to coarsest mesh resolution.
For each mesh, three bars are shown:  
    the recovered weld force in light gray, the adjusted
    weld force in darker gray, and the target value in black.
The optimal temperature for the medium weld speed was \valUnit{1470}{K};
    for the fast weld speed, it was \valUnit{1530}{K}.
Note that the optimal temperatures for the finest mesh were the
    same as those found for the coarsest mesh.
Thus, by using the adjusted weld force, 
   we found consistent probe temperatures for all three weld speeds.
%
%
%

%
%
%

As a final check, we compared the power usage 
   with the empirical data.
The empirical data gives a constant torque of \valUnit{47}{Nm}
   and a constant rotation speed of \valUnit{225}{rpm}
   for all three weld speeds.
Those values correspond to a constant  energy input from 
   the tool rotation of \Watts{1118}; a much smaller energy input
   comes from the linear motion of the tool.
The information we do not have, however, is how much 
   energy is removed by cooling of the tool.
In our study,  higher tool temperatures
   were required for the simulation weld force to match
   the experimental weld force at higher weld speeds.
Specifying a higher probe temperature in the simulation results in
  higher flux across the probe-workpiece interface and  greater
  heat input into the workpiece.
However, more material moves past the probe with higher weld
  speed, resulting in this application with the heat input per unit length
  of weld  decreasing with increasing weld speed.  
Nevertheless, the fraction of the total heat input being deposited 
  in the workpiece increases with increasing weld speed, approaching
  the entire input for the highest weld speed of 4 ipm.  
The power usage, including both frictional heating 
    and mechanical deformation heating,  is shown in Table~\ref{tab:power}.
All these numbers fall well within the empirical range.
The slower cases indicate significant heat loss to 
    the environment or through cooling.
The energy per unit length is generally of interest 
    and is also given in the table.
%
%
\begin{table}
    \caption{Power Usage}
    \label{tab:power}
    \centering
    \begin{tabular}{||c|c|c||} \hline
	\textbf{Weld Speed} & \textbf{Probe Heating Rate}  
                       & \textbf{Heat per Weld Length} \\
	\textbf{(in/min)} & \textbf{(W)}  
                       & \textbf{(J/mm)} \\\hline
            1.0  &  396   &  966 \\ \hline
            2.5  &  696   &  694 \\ \hline
            4.0  &  977   &  615 \\ \hline
    \end{tabular}
\end{table}

\subsection{Influence of Torque}
\label{sec:applic.parm_study_torque}
 
\begin{figure}[htpb]
  \begin{center}
    \includegraphics[width=4.0in]{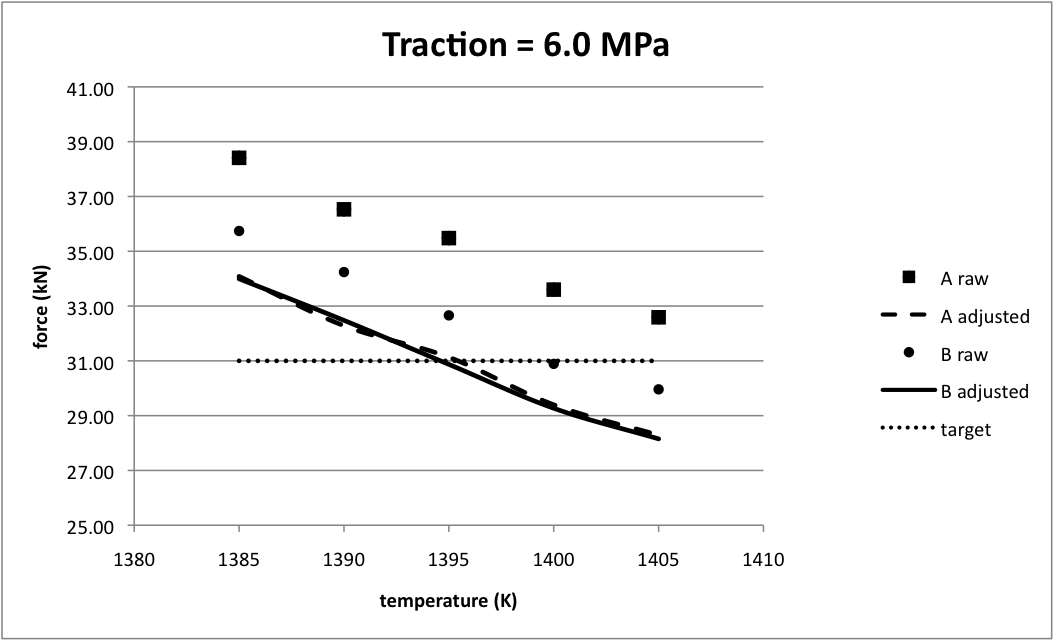}
    \includegraphics[width=4.0in]{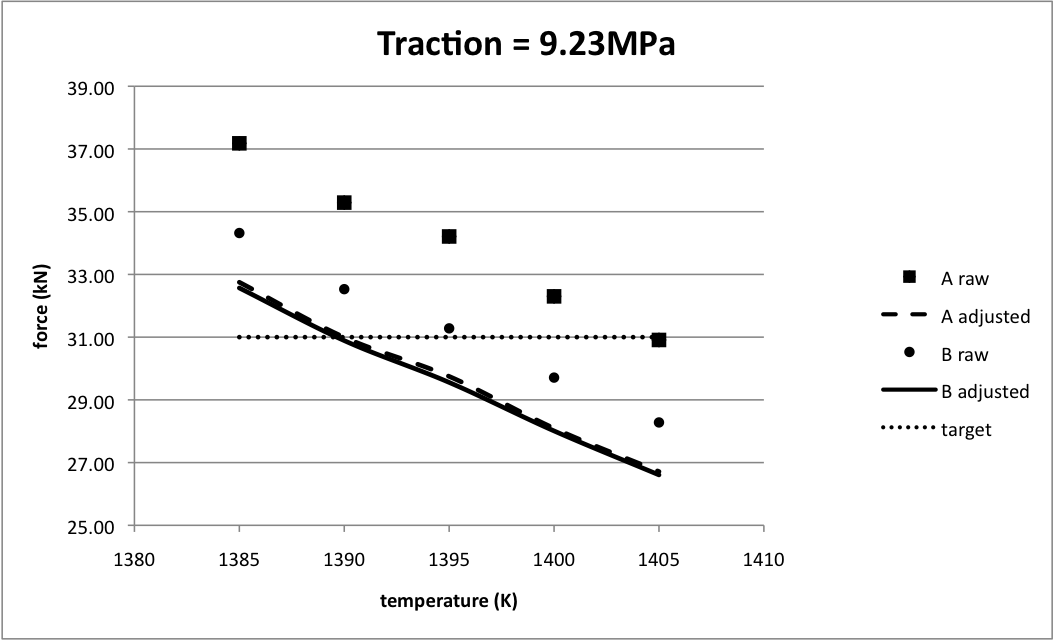}
    \includegraphics[width=4.0in]{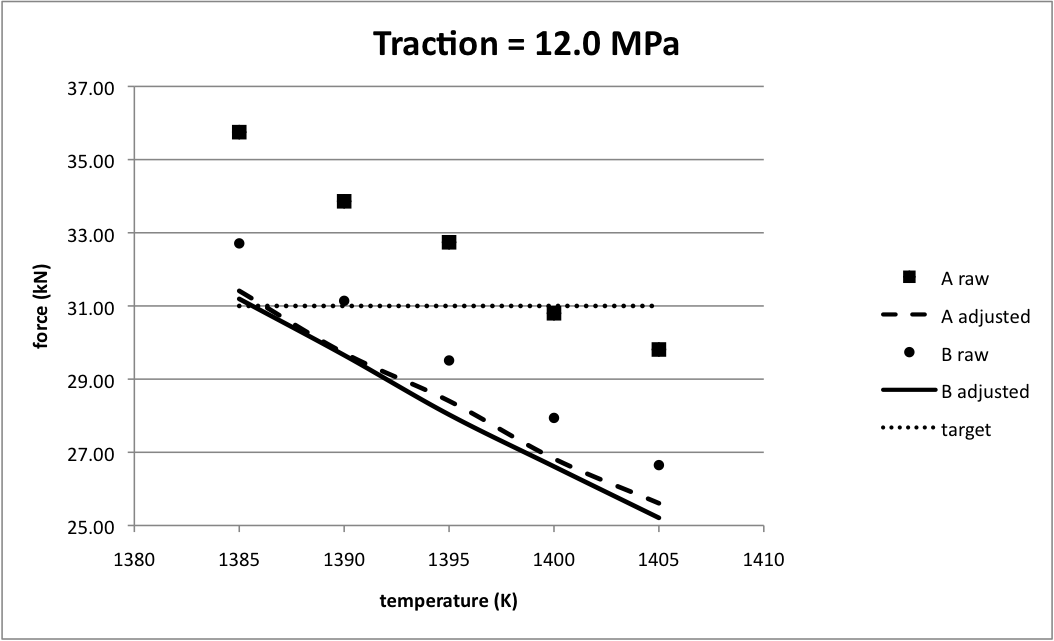}
    \caption{Weld force and temperature for two meshes and three values of torque.}
    \label{fig:torque-study}
  \end{center}
\end{figure}

As stated in Section~\ref{sec:applic.data}, the data set includes only an estimate
of the nominal torque, which is the same for all speeds.      
To study the sensitivity of the probe temperature to torque variations, 
    we repeated the suite of simulations for two values 
    of applied traction other than the empirical value of 9.23\,MPa.  
Values of 6.0\,MPa and 12.0\,MPa were assumed, which were estimated
   from experience to bracket the actual values of torque. 
The results for the slow weld speed are shown in Figure~\ref{fig:torque-study}.
Three plots are presented, one for each value of applied traction.
Each plot shows weld forces as a function of five temperatures ranging from
    1385\,K to 1405\,K.  
For each temperature, four values are shown:  the
    recovered weld force and the adjusted weld force, based on torque error,
    for both Mesh~A and Mesh~B.
The adjusted weld forces align  
    to within 1\% in all three cases.  
Also shown in each plot is a horizontal line indicating the empirical
    weld force of 31\,kN.
The probe temperature is taken to be where the target line intersects 
    the line of adjusted weld force.
%
%
The sensitivity of the probe temperature to changes in torque is small.  
For the slow weld speed, as the applied torque doubles, 
    the probe temperature required to produce
    the same weld force varies by only 10\,K, going from 1395\,K down to 1385\,K.
The two cases of faster weld speeds produced very similar results.

%
%
\section{Field Quantities}
\label{sec:discussion}
In FSW, friction between the probe and the workpiece acts to heat and shear the material in the 
vicinity of the probe.  The elevation of temperature upstream of the probe lowers the
flow stress and makes it possible to advance the probe with reasonably low forces.  
With the proper welding parameters, the ductility is adequate to accommodate the
large shear strains without inducing unacceptable levels of damage.
The simulations provide coupled solutions to the mechanical, thermal, and state evolution
problems, as described in Section~\ref{sec:sim}.   Each of these problems involves aspects
of the material's response during FSW that bears on the quality of the weld.  
Having generated solutions that match the empirical data and that are adequately converged, 
we can examine those solutions for insight into how the process deforms, heats
and alters the material.   

\subsection{Thermomechanical response}
\label{subsec:mech_response}

\begin{figure}[htpb]
  \begin{center}
    \includegraphics[width=2.5in]{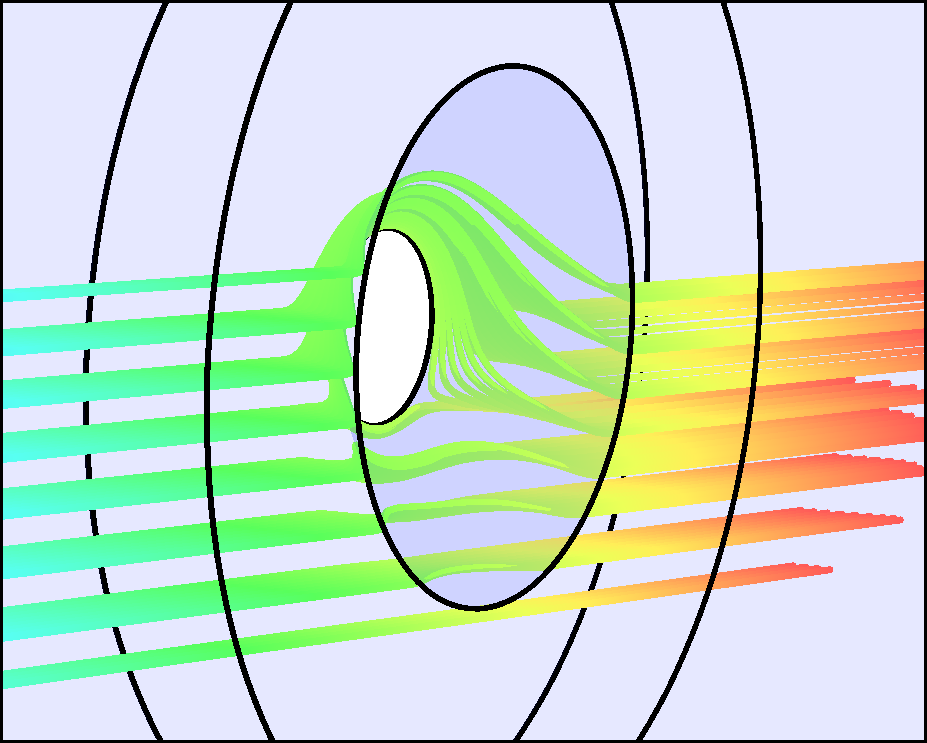}
    \includegraphics[width=2.5in]{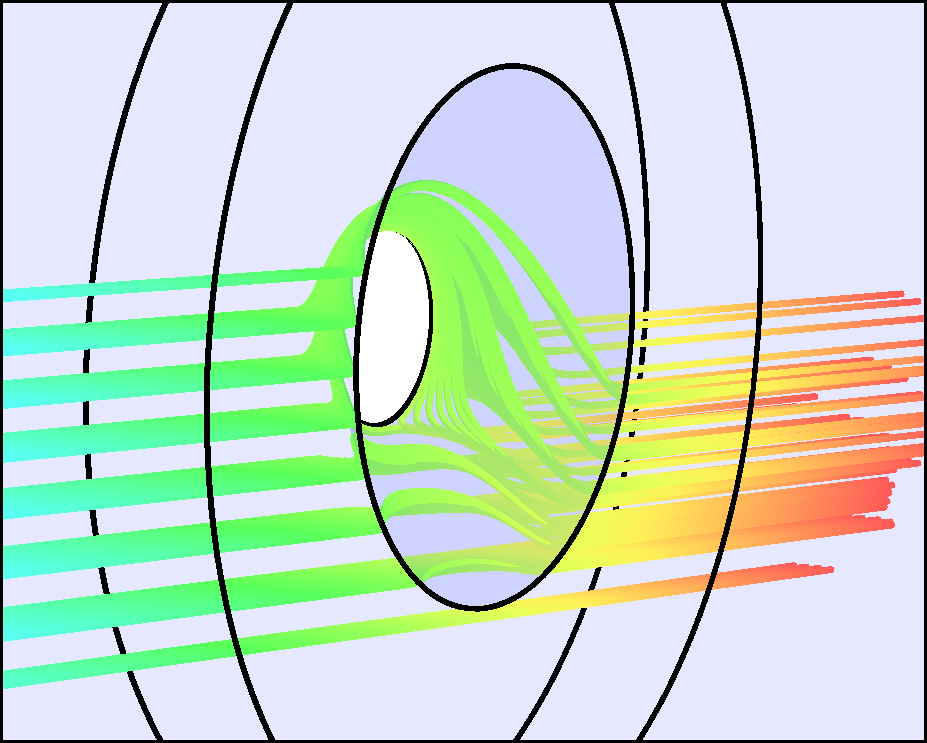}
    \includegraphics[width=2.5in]{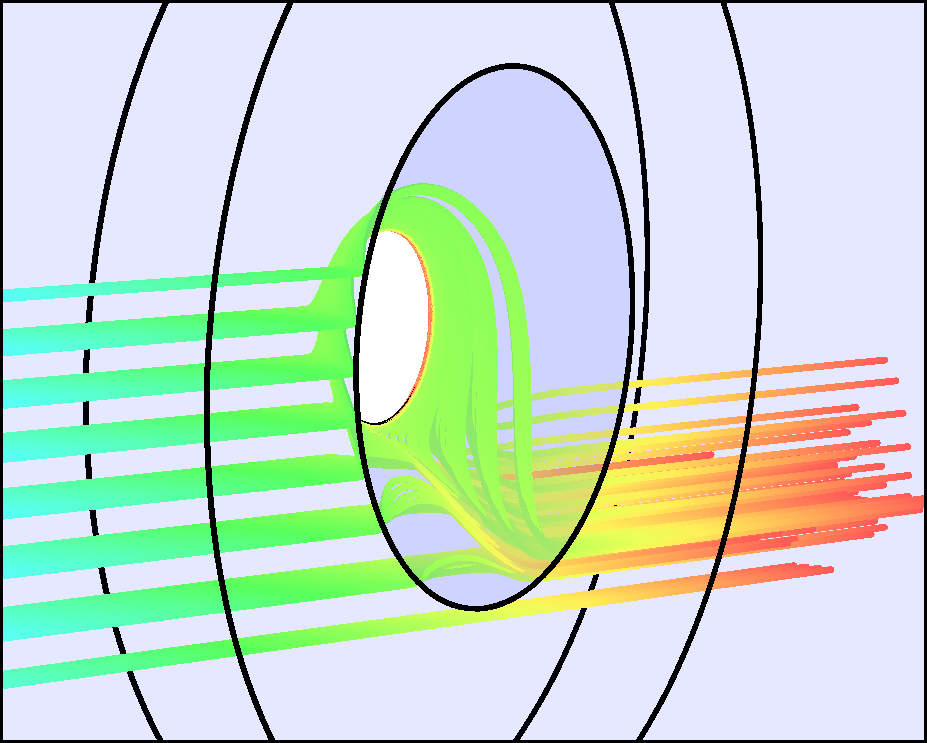}
    \caption{Streamlines illustrating flow pattern:  slow and medium on top, fast on bottom.
Color shows residence time.}
    \label{fig:streamlines}
  \end{center}
\end{figure}

The primary variable of the mechanical problem is the velocity field.  
Material enters the control volume and flows past the probe.  
Material in the path of the probe is drawn around the probe preferentially in one direction by the frictional tractions of the rotating probe.  
This occurs in a relatively thin deformation zone in close proximity to the probe surface 
     as can be seen from streamlines of the flow field shown 
     in Figures~\ref{fig:streamlines}.  
It may also be observed that there is a point upstream of the probe on the advancing side 
    where the flow splits into the part that passes the probe on the  retreating side 
    and the part that passes the probe on the advancing side.  

The weld speed has a clear influence on the overall flow pattern.  
For the slow case, the streamlines are nearly symmetric ahead of
    and behind the probe.
As the weld speed increases, the material is pulled
   more strongly to the advancing side, and the 
   flow also becomes more three-dimensional.
For the fast case, there is a strong deformation zone
   behind the probe on the advancing side where the
   separated material is being rejoined.

%
%
%
%
%
\begin{figure}[htpb]
  \begin{center}

    \includegraphics[width=2.4in,angle=-90]{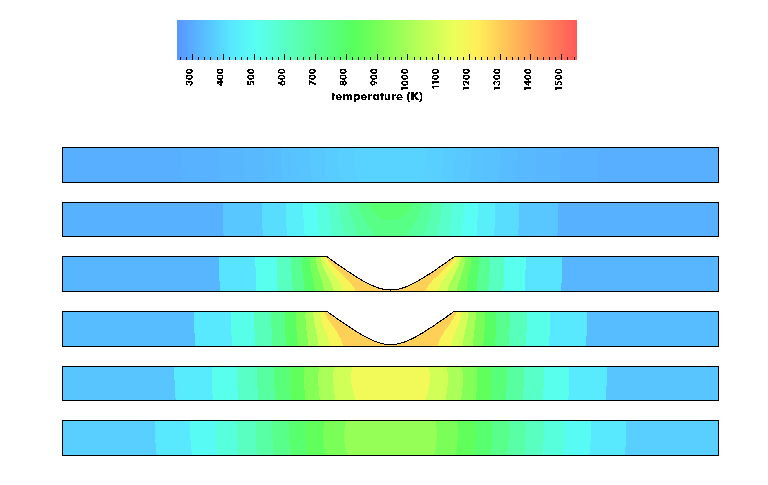}
    \includegraphics[width=2.4in,angle=-90]{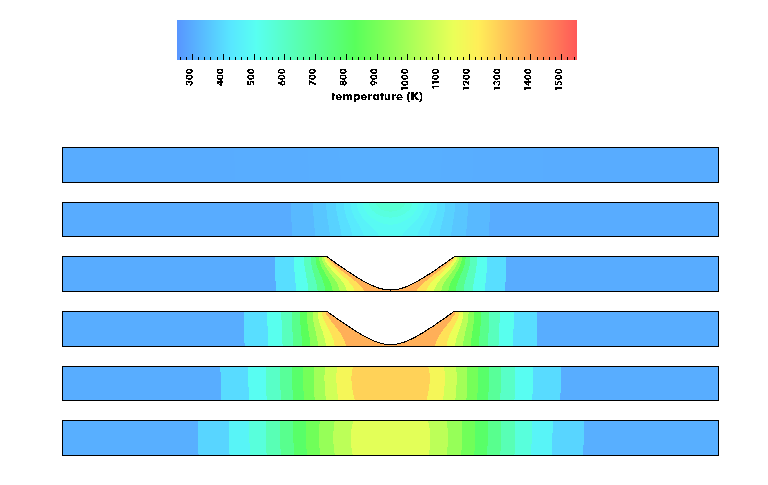}
    \includegraphics[width=2.4in,angle=-90]{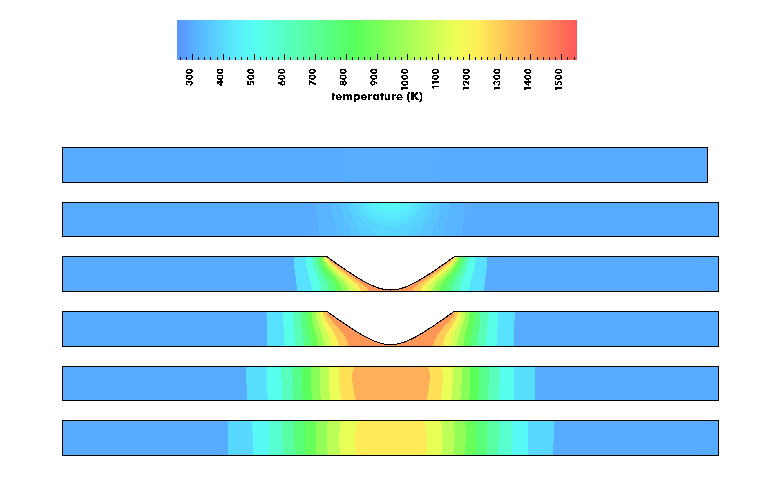}
           
    \includegraphics[width=2.4in,angle=-90]{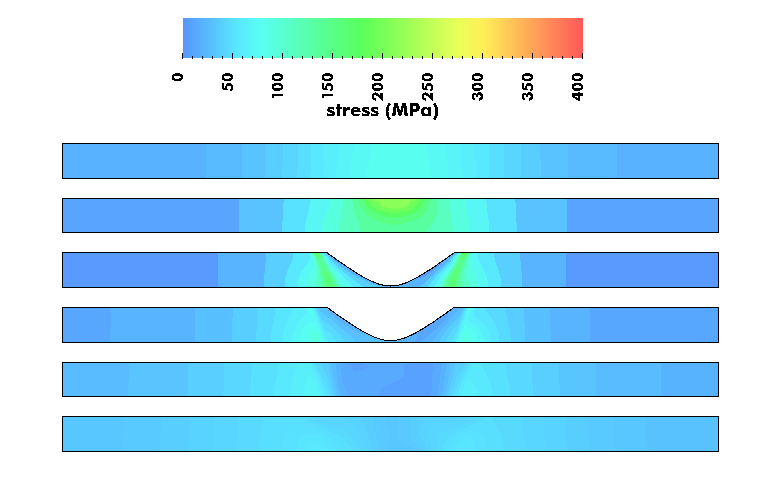}
    \includegraphics[width=2.4in,angle=-90]{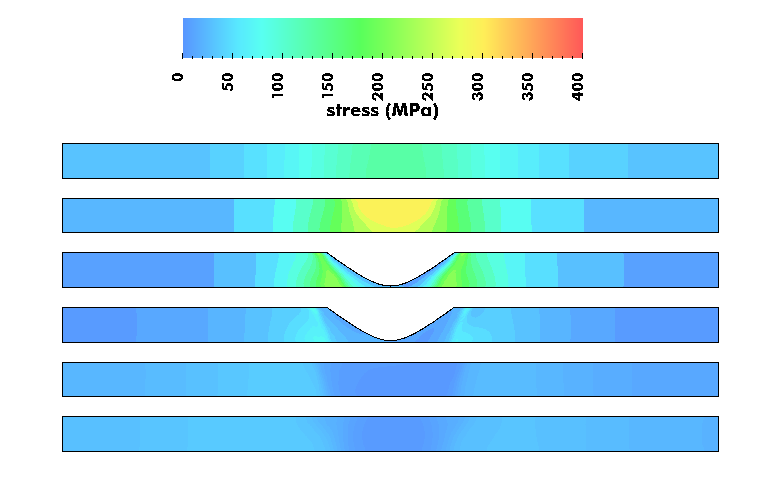}
    \includegraphics[width=2.4in,angle=-90]{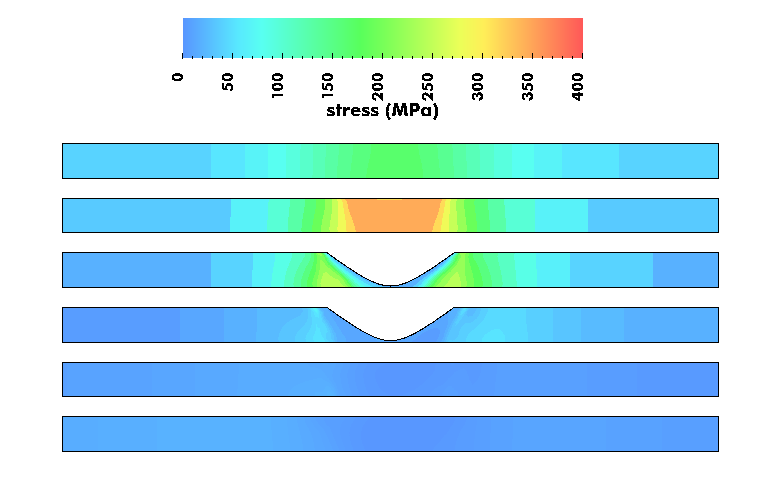}
                      
    \includegraphics[width=2.4in,angle=-90]{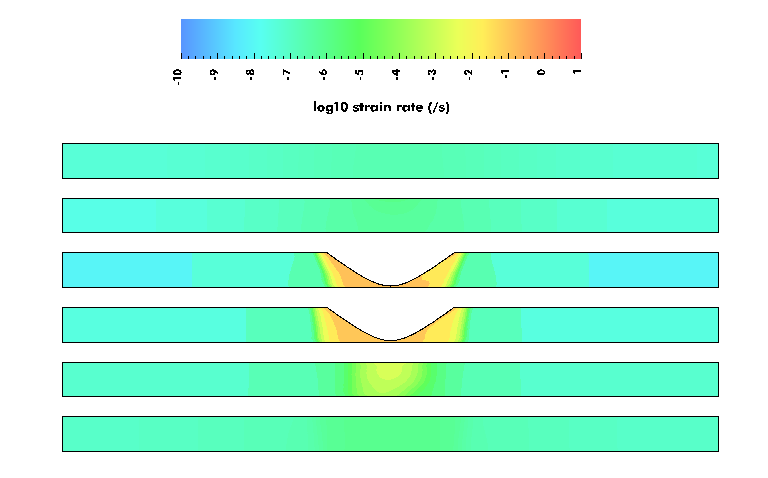}
    \includegraphics[width=2.4in,angle=-90]{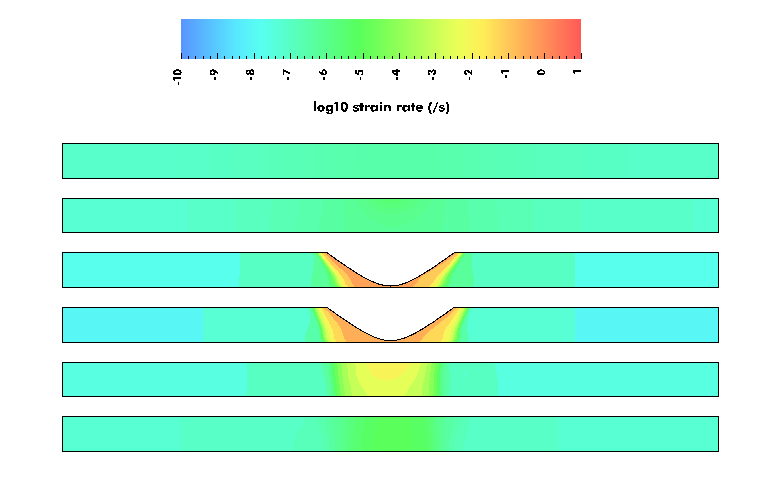}
    \includegraphics[width=2.4in,angle=-90]{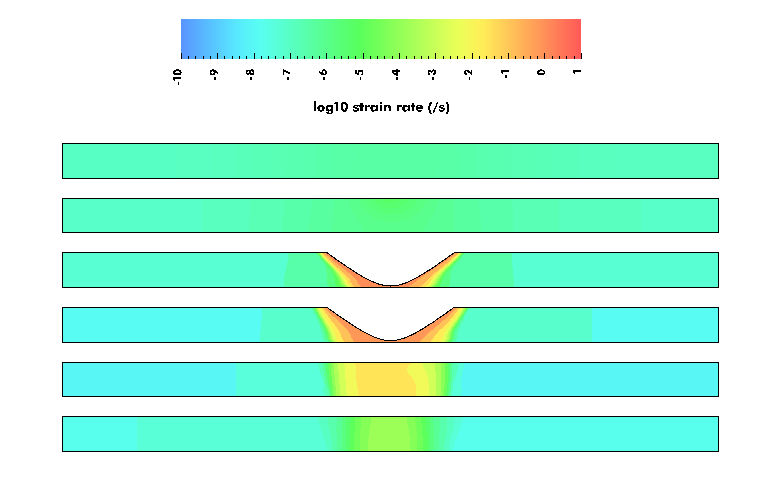}
    \caption{Temperature (top), effective stress (center), and strain rate(bottom, log scale) for 
    increasing weld speed  from left to right.  For each field, slices perpendicular to the weld direction are shown
    from the inlet on the right to the outlet on the left.}
    \label{fig:def-stress}
  \end{center}
\end{figure}

Because of the highly coupled nature of the temperature and velocity,
   we shown distributions of temperature, deformation rate and effective stress 
   together in Figure~\ref{fig:def-stress}.
The temperature profile is similar in shape for all three weld speeds, 
   but narrows with increasing weld speed as the material behind the probe
   spends less time near the heat source.
In general, the stress and deformation rate display some similarities, 
     but due to the nonlinear, temperature-dependent behavior 
     specified by Equation~\ref{eq:constitutive-law}, they are not directly proportional to each other.
Upstream of the probe, high values of the effective strain rate are limited to  a 
      narrow zone quite close to the probe surface.  
Downstream of the probe, the zone widens, reflecting the reduced rate sensitivity in 
Equation~\ref{eq:constitutive-law} at higher temperature.  
The effective stress is largest directly upstream of the probe. 
This is mainly a consequence of the probe pressing against 
      the material under the action of the welding force.  
The material must deform to flow around the probe, which 
      requires a higher level of stress upstream where material is just being heated.   
Downstream of the probe, the 
    effective stress is substantially lower than upstream for two reasons.  
First, the material is at elevated temperatures
    and, second, the strength has been reduced 
    (as is discussed in Section~\ref{subsec:evolution-of-state}).
Note that the applied frictional traction on the probe is substantially smaller 
    than the peak values of the effective stress.  
This indicates that the forces necessary to induce plastic flow of the workpiece 
      are coming primarily from the normal tractions along the probe, 
      which are associated with the weld force rather than
      the torque.  
%
%

\subsection{Evolution of the Strength}
\label{subsec:evolution-of-state}
\begin{figure}[htpb]
  \begin{center}
    \includegraphics[width=4.5in]{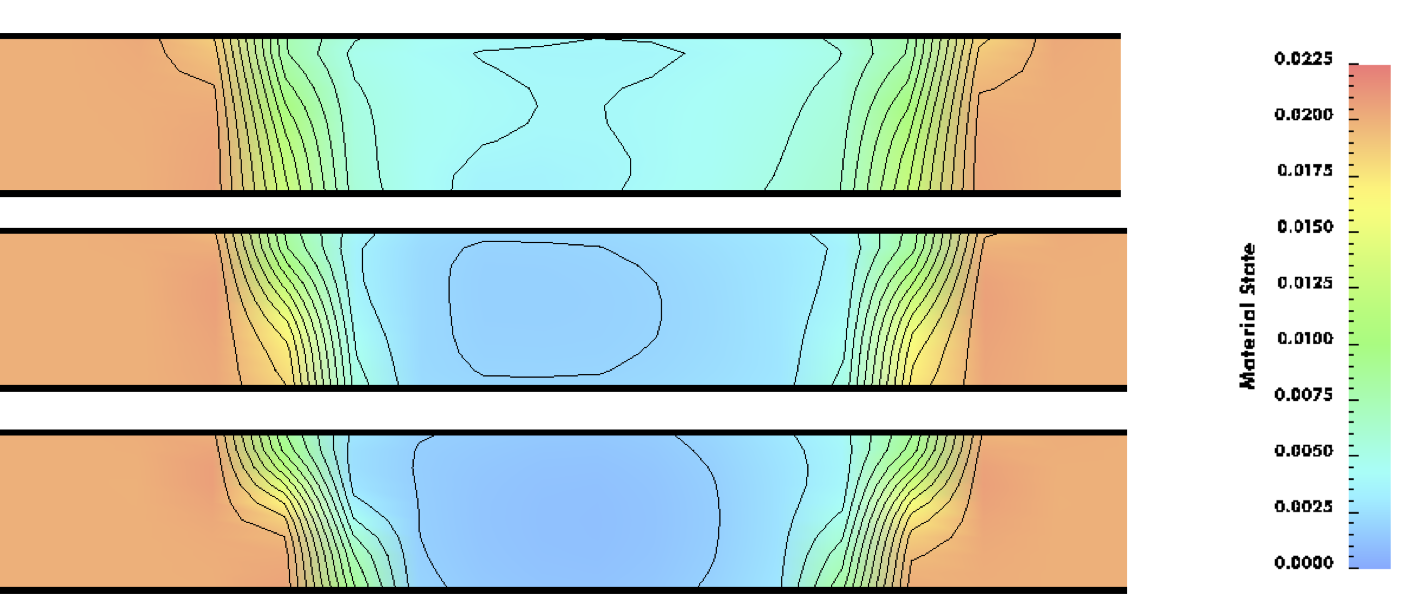}
    \caption{Material state cross-sections by weld speed (increasing from top to bottom).}
    \label{fig:state-x-sections}
  \end{center}
\end{figure}

Contour plots of the material state variable are shown in Figure~\ref{fig:state-x-sections}.
The material thermally softens immediately near the probe 
   quickly reaches a steady value dictated by the saturation stress.
The contour plots are qualitatively similar for 
   all weld speeds.
Cross sections on the outlet are shown in Figure~\ref{fig:state-x-sections}.
These images represent the material state after welding, but before possible thermal recovery.
The state variable has a large, softened region in the middle,
   which would empirically be called the weld nugget.
Bordering the nugget is a transition region characterized by a sharp gradient 
   in the state.  
Beyond the transition zone is base material 
   in which the state is unchanged. 
The contours in this transition zone sharpen with increasing weld speed,
   likely due to the narrowing band of heating behind the probe.
%
%

\subsection{Streamline Histories}
\label{subsec:streamline_histories}
\begin{figure}[htpb]
  \begin{center}
    \includegraphics[width=2.5in]{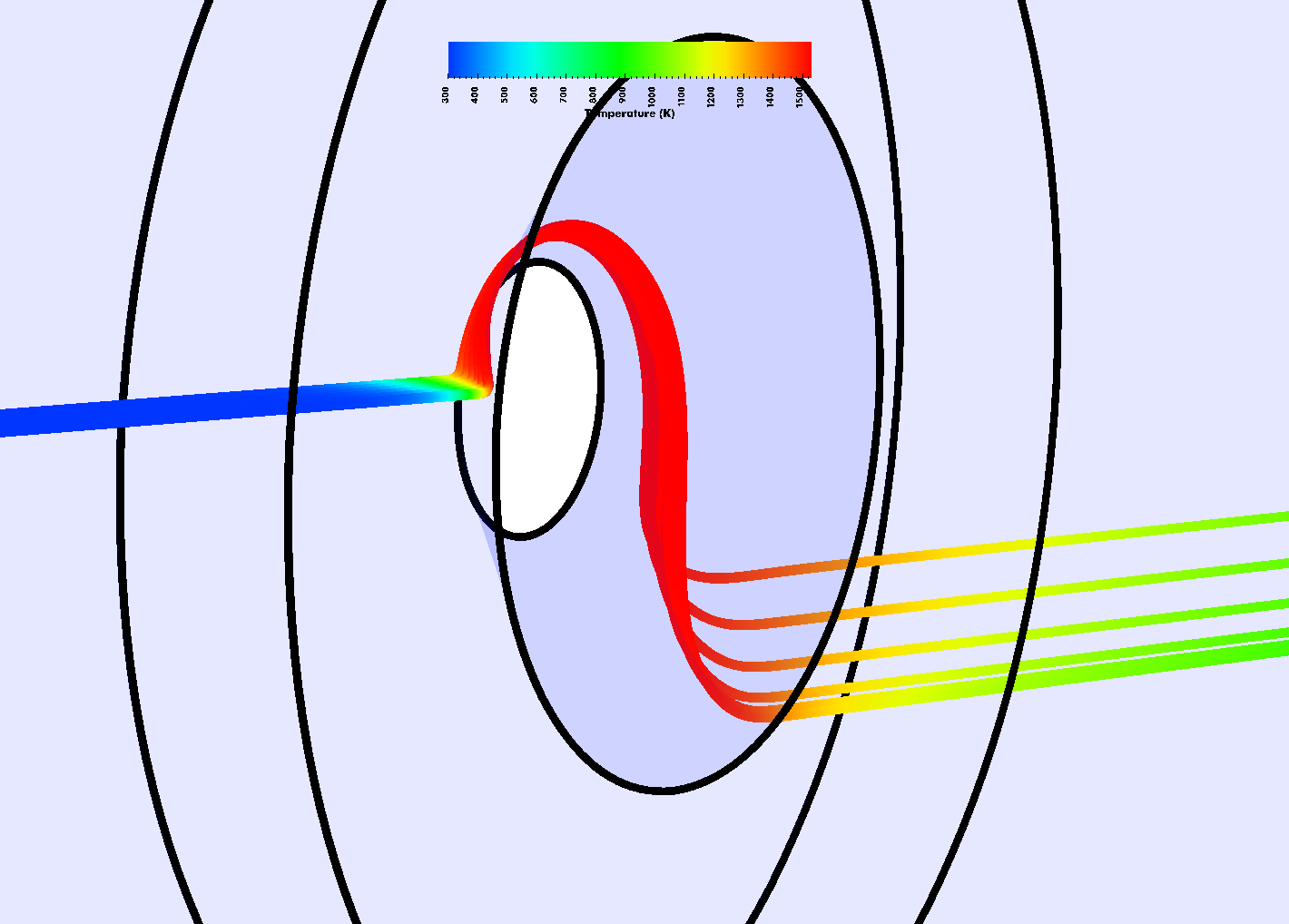}
    \includegraphics[width=2.5in]{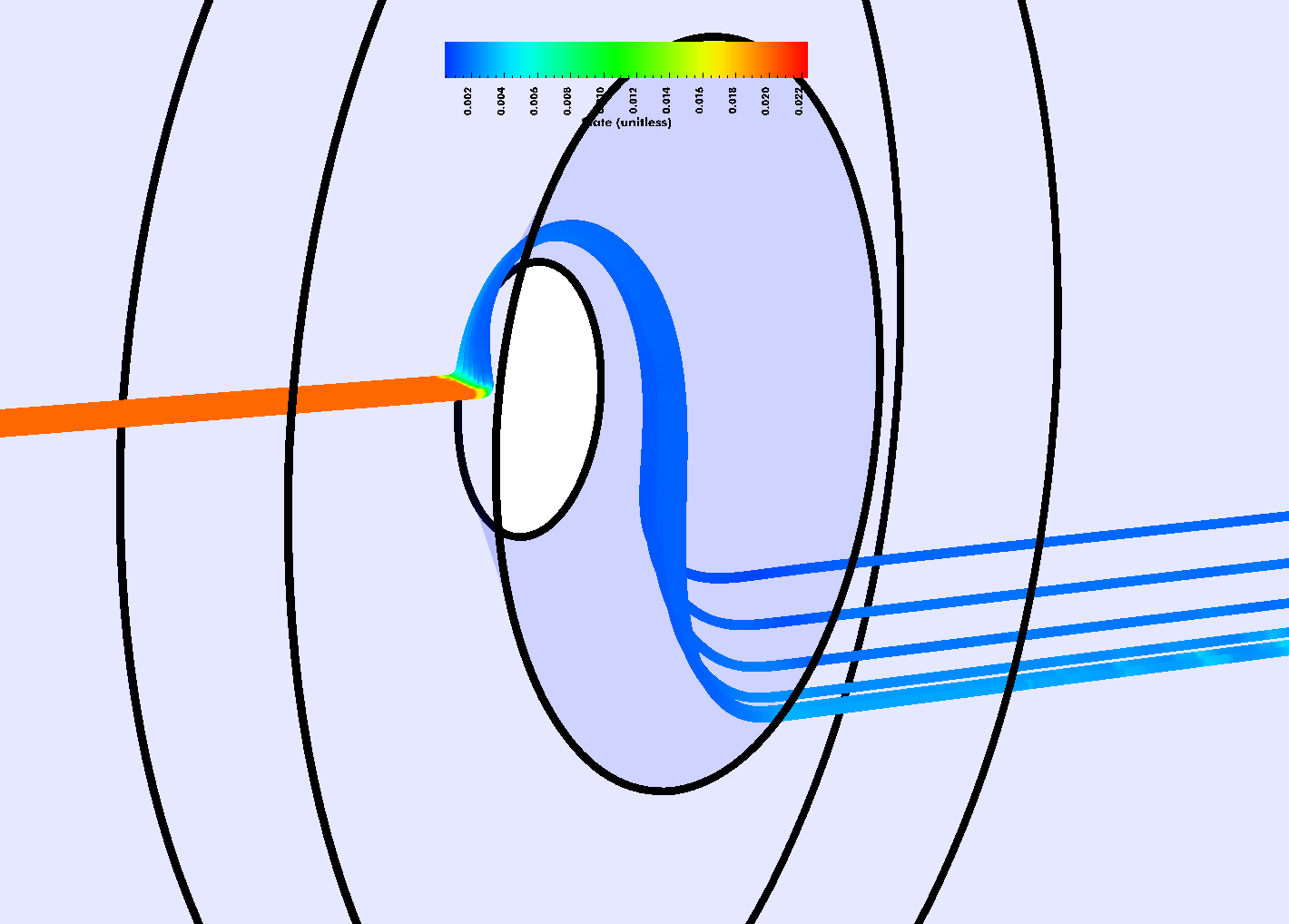}
    \includegraphics[width=2.5in]{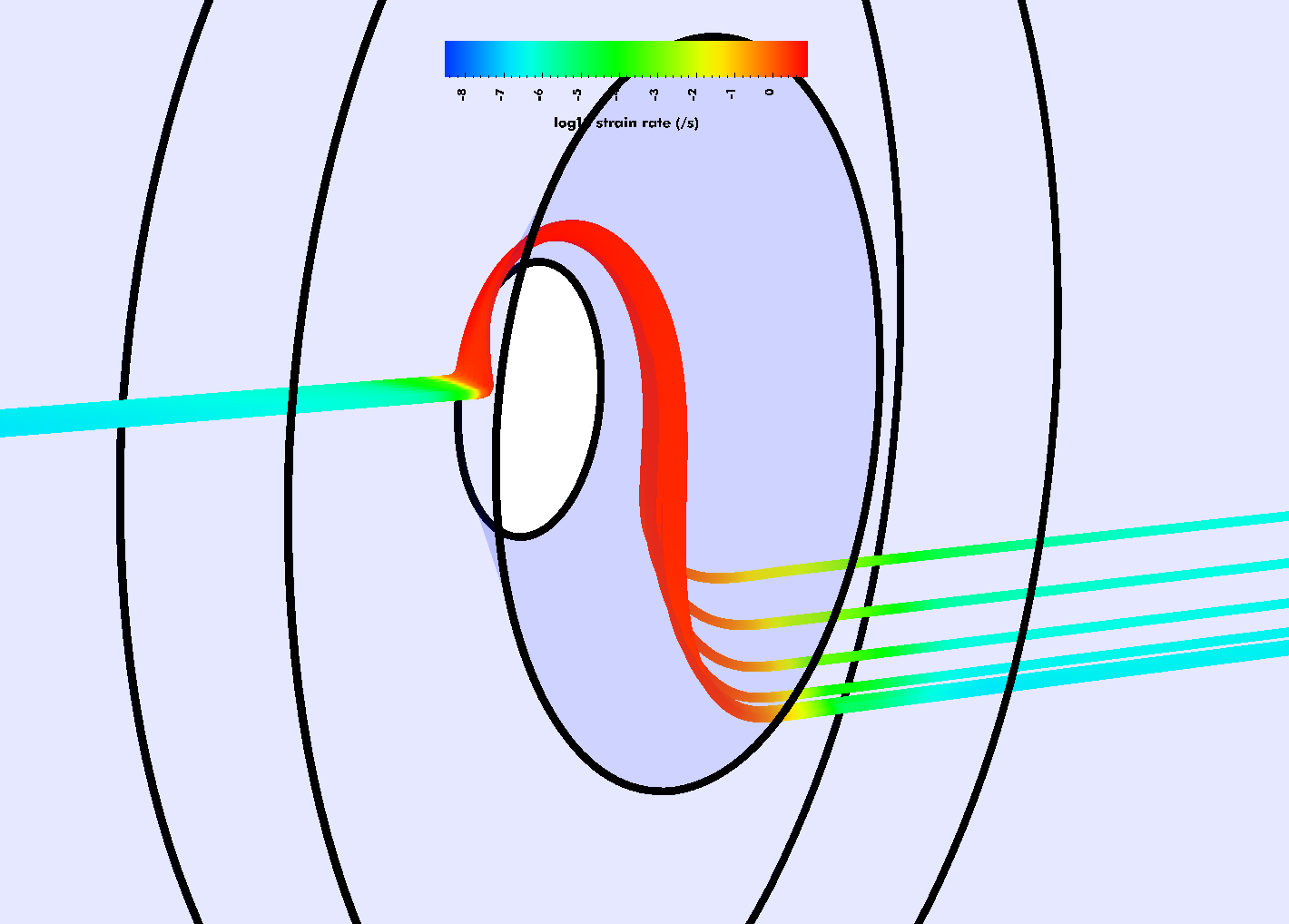}
    \includegraphics[width=2.5in]{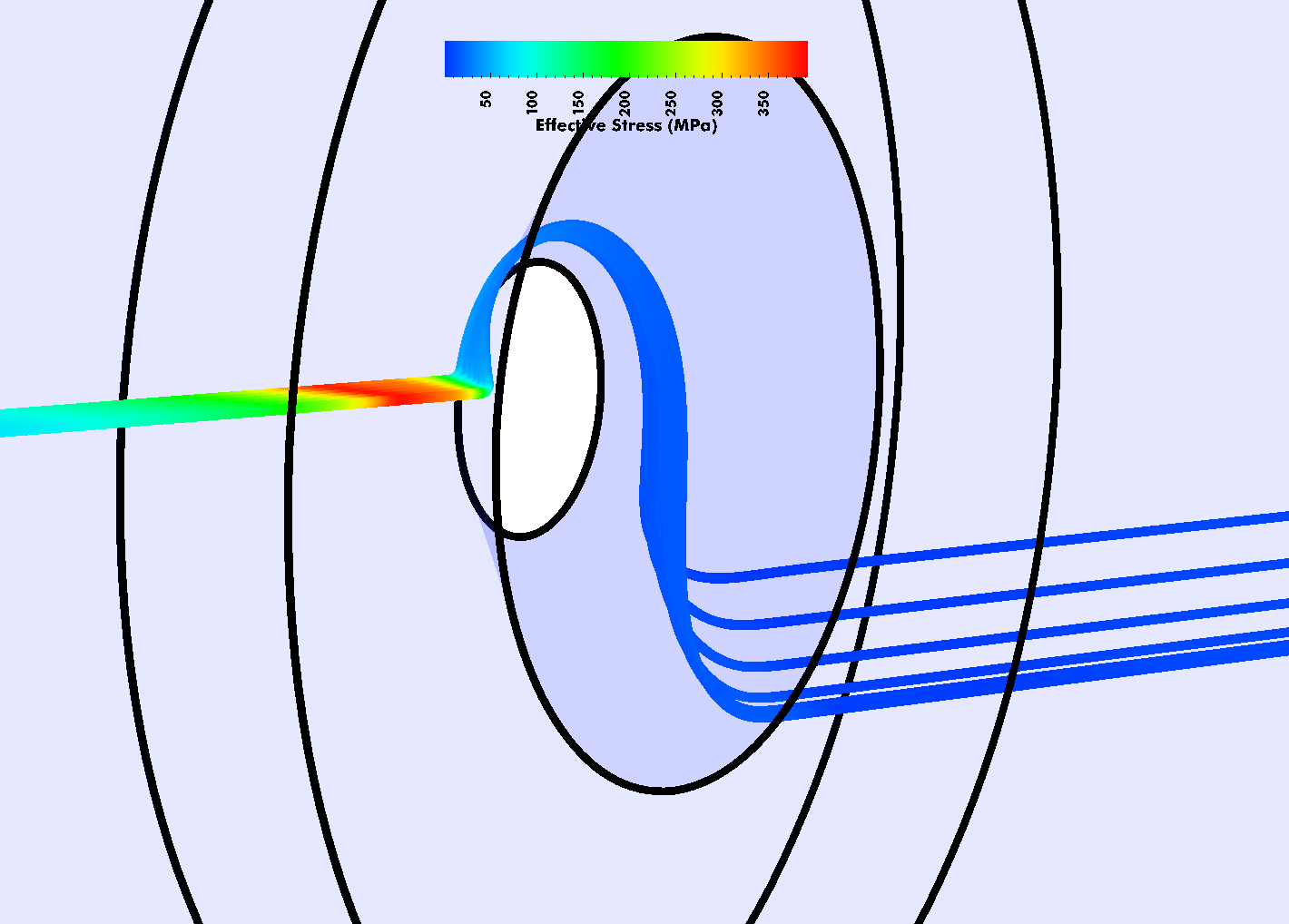}
    \caption{Data fields for fast case shown along streamlines. 
Top row:  temperature and state.  
Bottom row:  deformation rate and stress.}
    \label{fig:slhist}
  \end{center}
\end{figure}

The field plots presented in Figures~\ref{fig:def-stress} and \ref{fig:state-x-sections} give spatial distributions of 
	thermal and mechanical variables.    
It is often of interest to quantify the thermomechanical histories of material differential volumes (material points) throughout 
	the FSW process.  
Such information gives the processing path experienced by material points from different zones in the workpiece.   
These data can be used for direct comparison to embedded thermocouple records, 
	to verify if flow stress data encompasses the process conditions, 
	or as input to microstructural models, for example.   
Figure~\ref{fig:slhist} illustrates a collection of thermomechanical records for a set of material points
	originally along the centerline of the weld.    
One can see that the material is heated rapidly as it nears the probe.  
It begins intensive straining prior to being swept around the probe, 
	and continues to deform until it is just downstream of the probe.  
The action of the probe compresses the material in the weld direction while stretching it 
	in the transverse direction.
The combination of high temperature and deformation reduces the strength as material approaches the probe, 
	dropping the flow stress.
In contrast to the strain rate, the stress peaks further upstream of probe.  In this region, the influences of
thermal softening and strength recovery, has not yet reduced the flow stress as it does later in the process. 
%
%

\clearpage
%
%
\section{Summary}
\label{sec:summary}

%
%
%
%
%

In this work, we developed a methodology to determine
   values for temperature and traction on the probe interface
   in finite element simulations of friction stir welding.
The methodology goes beyond simple evaluation of 
   boundary values and includes verification of mesh 
   convergence and estimation of the error.
We applied the methodology to model an experiment 
   involving \TiFiveOne\ under three different weld speeds.
For each of the three cases, we calibrated our boundary values 
    using resultant weld forces and torques.
The simulation results were consistent with empirical power 
    measurements, the resulting data fields were discussed.
   
The boundary conditions are of the simplest form---we use
   a constant traction derived directly from the measured torque and a 
   constant temperature derived indirectly through its
   relation to weld force.
The traction magnitude is computed analytically from the 
   geometry of the probe surface, given the measured value
   torque.
The calculation of the probe temperature is indirect 
   and requires a few simulations.
Essential to the success of the method is a high quality
   material model that captures the material's thermal softening 
   with temperature.
The temperature on the probe interface is varied until the
   measured weld force is attained. 
In the end, the values of temperature and weld force 
   are consistent with measured resultants of torque and 
   weld force, and any other available data can be used 
   for validation.

We ran the simulations on multiple meshes to evaluate 
the discretization error and to assure that the solution was converged.
The recovered torque, the value obtained by postprocessing
   the simulation result, was compared with the analytically 
   expected value and used as an estimate of error.  
For the examples shown, the estimated errors decreased 
   as expected with the mesh size.
Furthermore, we were able to use the estimated error 
   in the torque to predict converged weld forces.
That allowed us to obtain consistent trends for  the probe temperature
   across all meshes.

Our simulations modeled friction stir welding of \TiFiveOne\ 
    at three weld speeds.
The power usage was within bounds dictated by the experimental conditions, 
    and the simulations indicated significant cooling
    through the tool.
After plotting the data fields, 
    a consistent picture of the stir welding process emerged.
The primary effect of the stirring is to heat the material
    ahead of the probe in order to soften it sufficiently 
    for the probe to advance though the material.
The peak stresses occur as the probe approaches, due to 
    the compressive action of the welding force on material
    that is not yet fully heated.
Once the material is drawn around the probe, it heats 
     rapidly.
The faster weld speeds showed greater lateral spread in the
    flow, with more vertical mixing.

\section*{Acknowledgments}
\label{sec:acknowledgments}

Support was provided  by the US Office of Naval Research (ONR) under contract N00014-09-1-0447. Thanks to Lars Wahlbin for useful discussions regarding {\it a posteriori} error estimation.   Thanks to Jennifer Wolk for sharing data on FSW of Ti5111.

\newpage
\bibliographystyle{unsrt}
\bibliography{fsw}

\begin{thebibliography}{10}

\bibitem{bely_book}
T.~Belytschko, W.~K. Liu, and B.~Moran.
\newblock {\em Nonlinear Finite Elements for Continua and Structures}.
\newblock John Wiley and Sons, 2000.

\bibitem{daw_84}
P.~R. Dawson.
\newblock A model for the hot or warm working of metals with special use of
  deformation mechanism maps.
\newblock {\em International Journal of Mechanical Sciences}, 26(4):227 -- 224,
  1984.

\bibitem{agr_daw_85}
A.~Agrawal and P.~R. Dawson.
\newblock A comparison of \protect{G}alerkin and streamline techniques for
  integrating strains from an \protect{E}ulerian flow field.
\newblock {\em International Journal for Numerical Methods in Engineering},
  21:853 -- 881, 1985.

\bibitem{daw_87}
P.~R. Dawson.
\newblock On modeling of mechanical property changes during flat rolling of
  aluminum.
\newblock {\em International Journal of Solids and Structures}, 23(7):947--968,
  1987.

\bibitem{cho_boy_daw_05}
Jae-Hyung Cho, Donald~E. Boyce, and Paul~R. Dawson.
\newblock Modeling strain hardening and texture evolution in friction stir
  welding of stainless steel.
\newblock {\em Materials Science and Engineering A}, 398:146--163, May 2005.

\bibitem{cho_boy_daw_2007}
J-H Cho, D~E Boyce, and P~R Dawson.
\newblock Modelling of strain hardening during friction stir welding of
  stainless steel.
\newblock {\em Modelling and Simulation in Materials Science and Engineering},
  15(5):469--486, 2007.

\bibitem{Liechty2008Modeling}
B.~Liechty and B.~Webb.
\newblock {Modeling the frictional boundary condition in friction stir
  welding}.
\newblock {\em International Journal of Machine Tools and Manufacture},
  48(12-13):1474--1485, October 2008.

\bibitem{Ainsworth1997Posteriori}
M.~Ainsworth and J.~Tinsley Odin.
\newblock {A posteriori error estimation in finite element analysis}.
\newblock {\em Computer Methods in Applied Mechanics and Engineering},
  142(1-2):1--88, March 1997.

\bibitem{Verfurth1996Review}
Rudiger Verfurth.
\newblock {\em {A Review of Posteriori Error Estimation \& Adaptive
  Mesh-Refinement Techniques}}.
\newblock John Wiley \& Sons, May 1996.

\bibitem{PETSc_2002}
S.~Balay, K.~Buschelman, W.~Gropp, D.~Kaushik, M.~Knepley, L.C. McInnes,
  B.~Smith, and H.~Zhang.
\newblock {\em PETSc Users Manual}.
\newblock Argonne National Lab., 2002.
\newblock (Argonne, IL, US).

\bibitem{kocks_mecking_2003}
U.~Kocks and H.~Mecking.
\newblock {Physics and phenomenology of strain hardening: the FCC case}.
\newblock {\em Progress in Materials Science}, 48(3):171--273, 2003.

\bibitem{engelman_1982}
M.S. Engelman, R.L. Sani, P.M. Gresho, and M.~Bercovier.
\newblock Consistent versus reduced integration penalty methods for
  incompressible media using several old and new elements.
\newblock {\em International Journal for Numerical Methods in Fluids},
  2:25--42, 1982.

\bibitem{lee_1989}
Y.S. Lee and P.R. Dawson.
\newblock Obtaining residual stresses in metal forming after neglecting
  elasticity on loading.
\newblock {\em Journal of Applied Mechanics}, 56(6):318--327, 1989.

\bibitem{book_fem2000}
T.~Belytschko, W.K. Liu, and B.~Moran.
\newblock {\em Nonlinear Finite Elements for Continua and Structures}.
\newblock JOHN WILEY \& SONS LTD, 2000.
\newblock chap. 7.

\bibitem{brooks:supg_1982}
A.~Brooks and T.~Hughes.
\newblock {Streamline upwind/Petrov-Galerkin formulations for convection
  dominated flows with particular emphasis on the incompressible Navier-Stokes
  equations}.
\newblock {\em Computer Methods in Applied Mechanics and Engineering},
  32(1-3):199--259, September 1982.

\bibitem{wolk:thesis}
Jennifer~N. Wolk.
\newblock {\em {Microstructural Evolution in Friction Stir Welding of Ti
  5111}}.
\newblock PhD thesis, University of Maryland, 2010.

\bibitem{Fonda2010Texture}
R.~W. Fonda and K.~E. Knipling.
\newblock {Texture development in near-{\OE}$\pm$ Ti friction stir welds}.
\newblock {\em Acta Materialia}, September 2010.

\bibitem{asm_handbook}
H.~E. Boyer and T.~L. Gall, editors.
\newblock {\em Metals Handbook}.
\newblock ASM International, 1985.

\bibitem{AtlasOfFormability}
P.~K. Chaudhury and D.~Zhao.
\newblock Atlas of formability: \protect{Ti-6Al-4V ELI}.
\newblock Technical report, National Center for Excellence in Metalworking
  Technology, 1992.

\bibitem{Pilchak2011Microstructure}
A.~Pilchak, W.~Tang, H.~Sahiner, A.~Reynolds, and J.~Williams.
\newblock {Microstructure Evolution during Friction Stir Welding of
  Mill-Annealed Ti-6Al-4V}.
\newblock {\em Metallurgical and Materials Transactions A}, 42(3):745--762,
  March 2011.

\end{thebibliography}
%
%
\end{document}